# Unique achondrite Northwest Africa 11042: Exploring the melting and breakup of the L Chondrite parent body

Zoltan Vaci[1,2], Carl B. Agee[1,2], Munir Humayun[3], Karen Ziegler[1,2], Yemane Asmerom[2], Victor Polyak[2], Henner Busemann[4], Daniela Krietsch[4], Matthew Heizler[5], Matthew E. Sanborn[6], and Qing-Zhu Yin[6]

[1]Institute of Meteoritics, University of New Mexico, Albuquerque, NM.  [2]Department of Earth and Planetary Sciences, University of New Mexico, Albuquerque, NM.  [3]National High Magnetic Field Laboratory and Dept. of Earth, Ocean & Atmospheric Science, Florida State University, Tallahassee, FL 32310, USA.  [4]Institute of Geochemistry and Petrology, ETH Zürich, Zurich, Switzerland.  [5]New Mexico Bureau of Geology, New Mexico Institute of Mining and Technology, Socorro, NM.  [6]Department of Earth and Planetary Sciences, University of California-Davis, Davis, CA.

## Abstract

Northwest Africa (NWA) 11042 is a heavily shocked achondrite with medium-grained cumulate textures. Its olivine and pyroxene compositions, oxygen isotopic composition, and chromium isotopic composition are consistent with L chondrites.  Sm-Nd dating of its primary phases shows a crystallization age of 4100±160 Ma.  Ar-Ar dating of its shocked mineral maskelynite reveals an age of 484.0±1.5 Ma.  This age coincides roughly with the breakup event of the L chondrite parent body evident in the shock ages of many L chondrites and the terrestrial record of fossil L chondritic chromite.  NWA 11042 shows large depletions in siderophile elements (<0.01×CI) suggestive of a complex igneous history involving extraction of a Fe-Ni-S liquid on the L chondrite parent body.  Due to its relatively young crystallization age, the heat source for such an igneous process is most likely impact.  Because its mineralogy, petrology, and O isotopes are similar to the ungrouped achondrite NWA 4284 (this work), the two meteorites are likely paired and derived from the same parent body.

1.0 Introduction

NWA 11042 is a meteorite that was initially characterized as a primitive achondrite due to its igneous texture and affinity to L chondrites. Primitive achondrites are texturally nonchondritic meteorites that bear affinities such as similar bulk mineralogy, chemical composition, or isotopic systematics, to chondritic precursors (Weisberg et al. 2006). A great deal of diversity exists within primitive achondrites, and no single characteristic defines the group. Some are thought to represent melts, partial melts, or melt residues from planetary bodies that underwent localized heating events such as early-stage partial differentiation or impact (Weisberg et al. 2006). These include grouped meteorites such as the acapulcoites, brachinites, lodranites, ureilites, winonaites, and the silicates in IAB-IIICD iron meteorites (Mittlefehldt et al. 1998). All of these groups plot either on or below the terrestrial fractionation line in triple oxygen isotope space, i.e. they have close to zero or negative $\Delta^{17}O$ values (Clayton and Mayeda 1996; Greenwood et al. 2017). In contrast to differentiated achondrites such as the SNC (shergottite-nakhlite-chassignites) or HED (howardite-eucrite-diogenite) groups, which are thought to have originated from Mars and 4 Vesta respectively (McSween 1985, 1994; Ruzicka et al. 1997; Treiman et al. 2000), the natures of the primitive achondrites' parent bodies remain controversial (e.g. Benedix et al. 2005; Warren 2011; Gardner-Vandy et al. 2013; Keil 2014; Dhaliwal et al. 2017; Goodrich et al. 2017; Greenwood et al. 2017; Keil and McCoy 2018). Because of their chemical affinity to chondritic material (Weisberg et al. 2006), they are thought to represent intermediate rocks within the evolutionary trend between primitive chondrites and fully differentiated achondrites.

While the internal heating of chondritic parent bodies by radioactive decay is generally thought to form primitive achondrites, impact heating and melting can also produce the igneous textures found in some achondrites. The partial to wholesale melting produced by impact is capable of erasing any previously chondritic texture just as completely as heating by radioactive decay. Additionally, large impacts on chondritic parent bodies sometimes produce melt pools extensive enough to differentiate the metal and silicate portions of initially chondritic material (Mittlefehldt and Lindstrom 2001; Benedix et al. 2008; Wu and Hsu 2019). In this study, we present the mineralogy and petrology, oxygen and chromium isotopic composition, major and trace element geochemistry, noble gas composition, and geochronology using $^{147}Sm/^{143}Nd$ and $^{40}Ar/^{39}Ar$ for NWA 11042. We also compare its petrology with that of the primitive achondrite

NWA 4284 and its geochemistry with that of the L-impact melt PAT 91501 (Benedix et al. 2008). Our results suggest that this newly discovered meteorite formed through impact melting rather than primary differentiation of the L chondrite parent body.

## 2.0 Samples and Methods

NWA 11042 was purchased in Morocco by Abdelhadi Aithiba in 2016, and a deposit sample resides at the University of New Mexico. Two small pieces were chipped off the main mass (90.1 g), mounted in epoxy, and polished to 0.05 μm smoothness. Two thin sections were also produced from additional sawed pieces. The thin sections were used for optical and electron microscopy, and the epoxy sections were analyzed using laser ablation inductively-coupled plasma mass spectrometry (LA-ICP-MS). Eight acid-washed fragments were measured for their oxygen isotopic composition. A chip of 1.5 g was crushed in a mortar and pestle and sieved to separate 500 μm, 250 μm, 150 μm, and ≤90 μm sized mineral grains. The ≤90 μm portion was ground into powder as it likely represented close to the bulk composition of the meteorite. The other portions were immersed in water and sonicated with a Fisher Sonic Dismembrator Model 300 to remove any impurities from individual grains. These grains were then magnetically separated, and then olivine, pyroxene, maskelynite, and melt pocket groups were hand-picked. Large mineral grains were selected by hand, while grain aggregates were further ground and sonicated in order to recover individual minerals. Care was taken to avoid any weathered material, which is minimal as the fusion crust is intact and metal grains appear unaltered. The olivine, pyroxene, melt pocket, and whole rock groups were used for Sm-Nd radiometric dating. Some of the maskelynite and whole rock samples were analyzed for their noble gas composition, including Ar-Ar dating.

A thin section of NWA 4284 on loan from Cascadia Meteorite Laboratory was also analyzed by electron microprobe for this study. A thick section of PAT 91501, on loan from the National Meteorite Collection at the Smithsonian Institution, was also analyzed by LA-ICP-MS.

## 2.1 Oxygen Isotope Analysis

Eight fresh fragments of interior materials of NWA 11042 weighing between 1-2 mg were selected under a stereomicroscope to avoid any possible contamination from fusion crust. Oxygen isotopic analyses were performed using a $CO_2$ laser + $BrF_5$ fluorination system following modified procedures outlined in Sharp (1990). Oxygen isotope compositions were calculated using the following procedure:

The $\delta^{17,18}O$ values refer to the per mil deviation of a sample's ($^{17}O/^{16}O$) and ($^{18}O/^{16}O$) ratios from the V-SMOW standard values, respectively, expressed as $\delta^{17,18}O = [(^{17,18}O/^{16}O)_{sample}/(^{17,18}O/^{16}O)_{V\text{-}SMOW} -1]*10^3$. The delta values were then converted to logarithmic $\delta'$ in which $\delta^{17,18}O' = \ln(\delta^{17,18}O/1000 + 1)*1000$. The $\Delta^{17}O'$ values were obtained from the $\delta'$ values following $\Delta^{17}O' = \delta^{17}O' - 0.528 * \delta^{18}O'$. Typical analytical precision of the laser-fluorination technique is better than ± 0.02‰ for $\Delta^{17}O'$.

## 2.2 Cr Isotope Measurements

Chromium isotopic measurements were made on a small fusion crust-free chip of NWA 11042. A homogenized powder was generated from the aliquot after crushing in an agate mortar and pestle. A 17.68 mg aliquot of the homogenized powder was placed into a PTFE capsule along with a 3:1 mixture of $HF:HNO_3$ and sealed in a Parr pressure vessel stainless steel jacket. The sample was heated in an oven at 190° C for 96 h. After complete dissolution, including any refractory phases such as chromite, the sample solution was dried down and re-dissolved in 6 N HCl for separation of Cr through column chromatography. Chromium separation followed the previously described methods of Yamakawa et al. (2009). The isotopic composition of the purified Cr fraction was measured using a Thermo *Triton Plus* thermal ionization mass spectrometer at the University of California-Davis. Mass spectrometry and mass fractionation correction procedures followed those described in Sanborn et al. (2019). All isotope ratios are expressed as parts per 10,000 deviation (ε-notation) from the measured NIST SRM 979 Cr standard.

## 2.3 Electron and X-Ray Microanalysis

The thin sections and polished mounts were analyzed by electron probe microanalysis (EPMA) using a JEOL 8200 SuperProbe at UNM. Analytical techniques included qualitative energy-dispersive spectroscopy (EDS) and quantitative wavelength-dispersive spectroscopy (WDS), including WDS mapping. An accelerating voltage of 15 kV, a beam current of 20 nA, and a spot size of 1 μm were used for analyses. Currents were reduced to 10 nA and spot sizes were increased to 10 μm for phases that included volatile elements. Standards for WDS analysis were manufactured by the C.M. Taylor Standard Corporation in Sunnyvale, CA. EDS maps were also created using an FEI Quanta 3D field-emission gun (FEG) scanning electron microscope (SEM). Certain areas of these sections and a whole rock powder were scanned using a Rigaku D/Max Rapid II X-ray diffractometer with a Cu tube Kα source.

## 2.4 LA-ICP-MS Measurements

Laser ablation ICP-MS (LA-ICP-MS) measurements were performed on epoxy-mounted sections using an Electro Scientific Instruments New Wave™ UP193FX excimer laser ablation system coupled to a Thermo Element XR™ at the Plasma Analytical Facility, Florida State University. Analytical methods employed followed Yang et al. (2015) and Oulton et al. (2016). Mineral grains were analyzed with laser spot sizes of 50 or 100 microns, with a repetition rate of 50 Hz and a dwell time of 10 seconds of ablation. The bulk composition was determined by rastering the laser over an area of 8.6 mm$^2$ using a 150 μm spot size scanned at 25 μm/s with a repetition rate of 50 Hz (about 60 μm depth). Peaks analyzed, interference corrections, and standardization followed Yang et al. (2015, 2018). Major elements were determined on the same spots using methods described in Humayun et al. (2010). The USGS glasses BHVO-2g, BCR-2g and BIR-1g, NIST SRM 610, NIST 1263 a steel, and the iron meteorites North Chile (Filomena) and Hoba were used as standards to calibrate the abundances of about 60 elements (Yang et al. 2018). Detection limits were determined from the 3-sigma standard deviation of the blanks taken with the laser off. Peaks not detected are shown as n.d. Precision of LA-ICP-MS analyses for major elements on spots is in the 2-5% range (Humayun et al. 2010), and for trace elements it varies as a function of the concentration from 2-10%, except for elements near the detection limit (Yang et al. 2018).

## 2.5 Sm-Nd Dating

$^{147}$Sm/$^{143}$Nd analyses were conducted at the Department of Earth and Planetary Sciences at the University of New Mexico in Albuquerque, NM. The olivine, pyroxene, chromite/melt pocket, and whole rock mineral separates were ground into fine powders and dissolved in Teflon bombs with HF and HNO$_3$ for 72 hours, along with a blank sample. The solutions were transferred to 30 mL Teflon beakers and merged with a Nd/Sm mixed spike. This mixture was dried down and redissolved using 15N HNO$_3$ until a brown crust was obtained. The brown crust

was totally dissolved in several steps with 6N HCl, 2N HCl, 1N HCl, and finally 1N $HNO_3$, to equilibrate the spike and the sample and prepare the solution for loading into the first set of columns. These solutions were then run through two sets of 0.2 mL columns to separate out the Sr using Eichrom Sr spec resin and rare earth elements (REE) using Eichrom TruSpec resin. The REE solutions were dried, dissolved in 0.18N HCl, and run through 10 mL Eichrom Ln resin columns to separate the Sm and Nd. The Nd and Sm isotopic compositions of these cuts were measured using a Thermo Neptune™ multicollector inductively coupled plasma mass spectrometer. The La Jolla Nd standard and a spiked Sm elemental standard were measured with the sample runs. Data were reduced using an in-house spreadsheet. Sm-Nd isochron age and plot were generated using IsoPlot (Ludwig 2008).

## 2.6 Noble Gas Measurements

Two bulk samples, a small and a large one, referred to as NWA 11042 "S" and "L", respectively (~20 and ~100 mg, Table 3), and three grains of maskelynite (~100 μg each, Table 3) were analyzed for their noble gas compositions at ETH Zürich to constrain exposure and radiogenic gas retention history and to trace potentially present trapped noble gases.

For the bulk measurements, fragments of NWA 11042 were wrapped into Al foil and subsequently heated to 110°C for seven days in vacuum to reduce terrestrial atmospheric contamination. The gas was extracted in one step by fusion in a Mo-crucible heated by electron bombardment to ~1700°C. Gas-free re-extractions at ~1750°C demonstrated complete extraction. The gases were separated into three fractions (He and Ne; Ar; Kr and Xe) and analyzed for all stable isotopes with a custom-built sector-field mass spectrometer equipped with a Baur-Signer ion source (Baur 1980), an electron multiplier operating in counting mode, and a Faraday cup for large gas amounts exceeding $1 \times 10^{-9}$ $cm^3$ STP (Riebe et al. 2017a). Blanks were Al foil measured with the same procedure as the samples and amounted typically to 900, 150, 40, 1, and <1 (in $10^{-12}$ $cm^3$ STP) for $^4He$, $^{20}Ne$, $^{36}Ar$, $^{84}Kr$, and $^{132}Xe$, respectively. Blank corrections were <1.5% for all He and Ne isotopes. Ar, Kr, and Xe blank corrections were more significant: up to 4/20% for all Ar isotopes, ~20/40 % for $^{84}Kr$, and ~15/45 % for $^{132}Xe$ in the large/small sample, respectively.

Initially unaware of the details of NWA 11042's parent body history, we analyzed maskelynite grains ("Mask1-3") in addition to bulk material. ince The probability to trace noble gases trapped from a potentially present atmosphere that could have been maintained if the parent body were sufficiently large is higher in quickly cooled glass. We used another in-house-built, high-sensitivity mass spectrometer that enables a precise measurement of He and Ne in small samples. An integrated molecular drag pump concentrates the noble gases into an ion source which increases the sensitivity by about two orders of magnitude compared to conventional mass spectrometers (Baur 1999). To release the gases (preheated in the storage volume to 110°C for five days), each grain was heated individually by a Nd:YAG infrared laser ($\lambda$ = 1064 nm) until melting (detailed procedure in Riebe et al. 2017b). Re-extractions were performed at slightly increased laser power to ensure full gas extraction and, if significant (only for Mask1), added to the main extraction step. Blank analyses (obtained by laser heating of an area on the Al sample holder without any nearby sample yielded <1% of the sample signals and were hence negligible for all He and Ne isotopes.

The measured noble gases were resolved into cosmogenic (cos), radiogenic (rad), and trapped (tr) components. Adopting that $^3$He is entirely cosmogenic, $(^4He/^3He)_{cos}$ ~5.65 (Wieler 2002), and $^4He_{tr}$ is negligible, we obtain ~25 % of the $^4$He in the bulk samples is radiogenic. In the maskelynite grains we observe larger variations from 25-80% $^4He_{rad}$. Neon in the bulk samples is also essentially purely cosmogenic ($^{20}Ne/^{22}Ne$ = 0.840-0.848 and $^{21}Ne/^{22}Ne$ = 0.895-0.908). In contrast, the maskelynite grains contain small amounts of $^{20}Ne_{tr}$ ($^{20}Ne/^{22}Ne$ = 1.018-1.302, which is, however, too close to the endmember $Ne_{cos}$ to resolve $Ne_{tr}$). We assume $Ne_{tr}$ to be air. The $^{36}Ar/^{38}Ar$ ratios (available only for the bulk samples) are slightly higher than the typical cosmogenic ratio, ~0.65 (Wieler 2002), which indicates a small amount of $Ar_{tr}$. We performed a two-component deconvolution between $(^{36}Ar/^{38}Ar)_{cos}$ of ~0.56-0.62, derived with the bulk chemical composition of NWA 11042 and a model for ordinary chondrite composition (Leya and Masarik 2009), and $(^{36}Ar/^{38}Ar)_{tr}$ of 5.31-5.36 (covering air and Q composition) to determine $^{38}Ar_{cos}$. $^{40}$Ar is likely entirely radiogenic, as the measured $^{40}Ar/^{36}Ar$ ratios and $^{40}$Ar concentrations are very similar, which would be very fortuitous if $^{40}$Ar from air would have been added to the fragments of very different masses. However, since we cannot fully exclude a contribution of atmospheric $^{40}$Ar the Ar retention ages obtained with the K-Ar method represent upper limits.

The production rates (Table 6) for cosmogenic $^3$He, $^{21}$Ne, and $^{38}$Ar used for the calculation of the cosmic ray exposure (CRE) ages are based on $^{22}Ne/^{21}Ne$ as a shielding indicator, the bulk chemical composition of NWA 11042 (Table 1), and the model for ordinary chondrite composition of the meteoroid (Leya and Masarik 2009). The uncertainty derived from the model is not included in the results but assumed to be 15-20%. In addition to this most reliable method to obtain precise CRE ages, we determined Kr-CRE ages for comparison with a model for ordinary chondrites (Eugster 1988). NWA 11042 shows Rb, Sr, Y, and Zr concentrations (Table 1), the most important target elements for the production of Kr, that are very similar to those found in L chondrites (Lodders and Fegley 1998). Hence, we preferred this approach to a model for production rates in achondrites (Eugster and Michel 1995). In Eugster (1988), production rates for chondrites are given as a function of the shielding-dependent ($^{22}$Ne/$^{21}$Ne)$_{cos}$ ratio and the chemical composition of the different chondrites (here we assume L-chondritic composition). The error of the production rate is estimated to be <20% (Eugster 1988). The concentrations in NWA 11042 of the target elements Ba and La for the production of Xe deviate from typical L chondrite composition, as ~16.4 ppm Ba is high compared to average Ba of 4.1 ppm in L chondrites (Lodders and Fegley 1998). However, this enrichment of Ba is likely related to terrestrial weathering. Hence, we applied the same approach as for Kr to determine the Xe CRE age. The uncertainty for the production rate is estimated to be 5% (1σ) (Eugster 1988). Nominal gas retention ages were determined for the U, Th-He, and K-Ar chronometers using radiogenic $^4$He and $^{40}$Ar (Table 3 and 4, respectively) and concentrations of U, Th, and K from Table 1. Argon retention ages obtained with the $^{40}$Ar-$^{39}$Ar technique are generally much more precise, hence, the additional age determination for conventional K-Ar rather allows a consistency check instead of delivering a high-quality value.

## 2.7 $^{40}$Ar/$^{39}$Ar Dating

Fragments of whole rock and a maskelynite mineral separate were analyzed by the $^{40}$Ar/$^{39}$Ar incremental heating method. The samples were irradiated at the TRIGA reactor in Denver, Colorado for 12 hours in the NM-290 package along with Fish Canyon Sanidine interlaboratory standard FC-2 with an assigned age of 28.201 Ma (Kuiper et al. 2008). Ages are calculated with a total $^{40}$K decay constant of 5.463×10$^{-10}$ a$^{-1}$ (Min et al. 2000). $^{40}$Ar/$^{39}$Ar analyses were conducted at the New Mexico Bureau of Geology, New Mexico Institute of Mining and Technology in Socorro, NM.

After irradiation (without Cd shielding), the NWA 11042 samples were evacuated in an all metal argon extraction line and baked for 4 hours at 150°C. The samples were step-heated using an 810 nm diode laser, and each step consisted of 40 seconds of heating and 60 seconds of gas cleanup. Each extracted gas step was cleaned with a SAES GP 50 getter operated at 1.6 A and exposed to a cold-finger set at -125°C. Following gettering the gas was analyzed for argon isotopes using a Thermoscientific Helix MC-plus mass spectrometer operated in static mode. The collector configuration had $^{40}$Ar, $^{39}$Ar, $^{38}$Ar, $^{37}$Ar and $^{36}$Ar on the H2, H1, AX, L1 and CDD detectors, respectively. All Faraday detectors are equipped with $1\times10^{12}$ Ohm resistors and the CDD is an ion counting detector with a dead time of 20 ns. All data acquisition was accomplished with NM Tech Pychron software, and data reduction used Mass Spec (v. 7.875) written by Al Deino at the Berkeley Geochronology Laboratory. Extraction line blank plus mass spectrometer background values are averages of numerous measurements interspersed with the unknown measurements. These values are 6.0±0.6, 1.00±0.12, 0.1±0.1, 0.10±0.05, 0.1500±0.0075, x $10^{-17}$ moles from masses 40, 39, 38, 37, and 36, respectively. Plateau ages represent the inverse variance weighted mean of the selected steps, the integrated age is the summation of all steps, and the error is calculated by quadratically combining the individual errors. The plateau is defined by contiguous heating steps that overlap at the 2σ confidence level. Errors are reported at 2σ. Other than blank $^{40}$Ar and reactor derived $^{40}$Ar from the irradiation of $^{40}$K all $^{40}$Ar is assumed to be radiogenic (that is the trapped initial $^{40}$Ar/$^{36}$Ar is zero).

## 3.0 Results
### 3.1 Mineralogy and Petrology

NWA 11042 is a rock that exhibits cumulate textures (Fig. 1). Its mineralogy consists of ~400-800 μm subhedral grains of olivine (34.5 vol%), ~300-600 μm grains of anhedral orthopyroxene (36.2%) and augite (10.2 vol%), smaller grains of maskelynite (16.9%), chromite (1.2 vol%), troilite (0.4 vol%), and FeNi metal (0.7 vol%). The FeNi metal is mostly kamacite with small amounts of taenite. Trace amounts of apatite were also identified. The olivine is uniform in composition without any signs of magmatic zoning (Fo$_{75.4\pm0.2}$, n=20). The pyroxene is bimodal and comprised of orthopyroxene (Fs$_{20.2\pm0.4}$ Wo$_{4.0\pm0.4}$, n=29) and augite (Fs$_{10.5\pm0.5}$ Wo$_{38.5\pm0.1.5}$, n=17). Any plagioclase that the rock might have had has been shocked to glassy albitic maskelynite (Ab$_{83.4\pm1.7}$ An$_{11.9\pm1.8}$ Or$_{4.8\pm0.4}$, n=10) which is surrounded by radiating fractures projecting into surrounding olivine or pyroxene grains. Chromite grains (wt.% Al$_2$O$_3$=5.0±3.3, Cr$_2$O$_3$=56.4±1.7, Mg=3.0±0.4, Mn=0.79±0.3, Fe=29.1±1.2, TiO$_2$=2.1±0.8, n=13) are similar in composition to those of ordinary chondrites in general (Johnson and Prinz 1991). All FeNi metal is found paired with the troilite. The calculated 2-pyroxene equilibrium temperature is 1111 ± 38 °C (Putirka 2008).

NWA 11042 contains large (~500-5000 μm) shock-melt pockets scattered throughout the groundmass. These are heterogeneously distributed, but melt fraction is estimated to be 10-15% cross-sectionally. The largest pockets contain high-pressure phases such as ringwoodite, identified optically as large blue grains (Fig. 2) and confirmed using X-ray diffraction (Fig. A1). Some of the olivines around the pocket appear to have been stained brown due to shock-induced strain such as in the chassignite NWA 2737 (Bläß et al. 2010). Other olivines show undulatory extinction under cross-polarized light (XPL). EDS identification suggests that the shock-melt pockets contain relict partially melted major phases such as olivine, pyroxene, and plagioclase, in addition to mesostasis. Melted areas are enriched in free metal and sulfide relative to the groundmass. Shock veins are also present in the sample, but the pockets are much more prevalent. In backscattered electron (BSE) images, the melt pockets have dark (low Z) rims which show exsolving metal particles (Fig. 3).

NWA 11042 is mineralogically and petrologically similar to ungrouped achondrite NWA 4284 (Fig. 4), which contains the same phases in different modal abundance: olivine (34.1 vol%, Fo$_{75.2\pm0.8}$, n=12), orthopyroxene (36.6 vol%, Fs$_{20.5\pm0.1}$ Wo$_{3.2\pm0.5}$, n=13), augite (6.0 vol%, Fs$_{9.5\pm0.0}$ Wo$_{40.9\pm0.2}$, n=2), plagioclase (21.5 vol%, Ab$_{82.0\pm2.1}$ An$_{12.0\pm1.6}$ Or$_{6.0\pm1.7}$), chromite (1.3 vol%), and troilite/kamacite (0.4 vol%). NWA 4284, unlike NWA 11042, does not contain shock-melt pockets, and the plagioclase it contains has only partially been transformed to maskelynite. The olivine and pyroxene compositions in both meteorites are largely similar to those of L chondrites (McSween et al. 1991).

### 3.2 Trace Element Geochemistry

Major and trace element concentrations determined by LA-ICP-MS for NWA 11042 are shown in Table 1. Similar data gathered for PAT 91501 is shown in Table 2. The bulk rock REE patterns of NWA 11042 and PAT 91501 are almost completely flat and ~2× CI chondritic (Fig. 5-6). The plagioclase and maskelynite show positive Eu anomalies, while the pyroxenes show complementary negative Eu anomalies. This is characteristic of a somewhat reducing magmatic environment in which some of the Eu was incorporated into the plagioclase in a divalent state. The melt pockets show smaller positive Eu anomalies, indicating plagioclase input. They also contain lower concentrations of REE than the bulk rock, and as such appear to have undersampled augite. The REE pattern of PAT 91501 (Fig. 6) is largely similar to that of NWA 11042 but with a slight negative Eu anomaly in the bulk rock data.

Siderophile element compositions for NWA 11042 are compared with bulk ordinary chondrite compositions in Fig. 7. The bulk composition of NWA 11042 exhibits large compatible siderophile element depletions relative to bulk ordinary chondrites. This is reflected in the lower modal abundances of metal and troilite found in the meteorite. The pattern of siderophile element depletion is similar to that in the L-impact melt PAT 91501 (Fig. 7).

### 3.3 Stable Isotopic Data

NWA 11042 has a $\Delta^{17}O$ value of +1.03‰ and plots in the domain of the L ordinary chondrites on a triple oxygen diagram (Fig 1). The oxygen isotopic composition is resolved from both H and LL chondrites. The eight analyses form a trend whose slope falls in between the slope of the terrestrial fractionation line (TFL) and the carbonaceous chondrite anhydrous mineral (CCAM) line (Clayton and Mayeda 1999). The oxygen isotopic composition of NWA 11042 does not plot anywhere near that of any grouped achondrites, but its O isotopes are similar to ungrouped achondrites NWA 4284, 5297, and 6698 (Fig. 8) (Greenwood et al. 2017). The Cr isotopic composition of NWA 11042 is $\varepsilon^{53}Cr = +0.22 \pm 0.04$ and $\varepsilon^{54}Cr = -0.29 \pm 0.08$. When plotted on a $\varepsilon^{54}Cr$ versus $\Delta^{17}O$ diagram (Fig. 9), the values for NWA 11042 plot within the compositional field defined by ordinary chondrites.

### 3.4 Sm-Nd Dating

The ratios $^{147}Sm/^{144}Nd$ and $^{143}Nd/^{144}Nd$ of the olivine, pyroxene, melt pocket, and whole rock samples were plotted against each other to assess if the mineral phases form an isochron (Fig. 10) with the linear equation $^{143}Nd/^{144}Nd = {}^{143}Nd/^{144}Nd_0 + {}^{147}Sm/^{144}Nd*(e^{\lambda t} - 1)$, where $^{143}Nd/^{144}Nd_0$ is the initial value at time t, and $\lambda$ is the decay constant of the alpha decay of $^{147}Sm$ to $^{143}Nd$, $6.539 \times 10^{-12}$ a$^{-1}$. The olivine, pyroxene, and whole rock points form a line with t =

4.10 ± 0.16 Ga. Since the melt pockets plot below this line, they were likely partially reset at a much later time than the crystallization age of the rest of the rock.

## 3.5 $^{40}$Ar/$^{39}$Ar Dating

$^{40}$Ar/$^{39}$Ar dating of NWA 11042 whole-rock and maskelynite yield similar age spectra that plateau for the first ~80-90% of $^{39}$Ar released, followed by a slight increase in age for the final heating steps (Fig. 11). The maskelynite plateau age is 484.0 ± 1.5 Ma, while the whole-rock produces a less precise plateau age of 478.5 ± 5.5 Ma (2σ) (Fig. 11). The maskelynite data are more precise than the whole-rock data due to the higher amount of K (0.77 vs 0.07 wt% $K_2O$) in the maskelynite. Since the maskelynite is the major carrier of K, the whole-rock is likely dominated by degassing of maskelynite and thus both plateau ages likely reflect the closure age of the maskelynite. Isochron plots of the maskelynite steps, corrected for cosmogenic $^{36}$Ar (see section 3.6), are shown in the appendix (Fig. A2). The slopes of these plots show ages of 481.5 ± 1.6 Ma and 482.4 ± 1.7 Ma, and the similarities between these and the plateau ages suggest minimal non-radiogenic Ar components within the maskelynite. As noted in the methods section, these ages are based on the assumption of a trapped $^{40}$Ar/$^{36}$Ar value of zero and thus represent the maximum age of NWA 11042. For instance, if we used a trapped value of 10 and 20 to simulate potential atmospheric Ar contamination, the maskelynite age is reduced from 484 Ma to 477 and 468 Ma, respectively. Incorporating these $^{40}$Ar/$^{36}$Ar values imparts significant internal discordance to the age spectrum and therefore we suggest that it is unlikely there is significant contamination with modern atmosphere and that the age of 484.0 ± 1.5 Ma is accurate.

## 3.6 Noble Gas Measurements
### 3.6.1 Concentrations and Isotopic Compositions

He and Ne concentrations and isotopic compositions for bulk samples and maskelynite grains are given in Table 3. The results are mostly consistent within uncertainties for the two bulk samples. Larger variations can be observed compared to and amongst the concentrations and isotopic compositions of the maskelynite samples. Their $^{21}$Ne/$^{22}$Ne ratios are slightly lower (indicating slightly larger relative contributions of trapped components) than for the bulk samples but still within the range of typical cosmogenic values (Wieler 2002). However, the

$^{20}$Ne/$^{22}$Ne ratios of the maskelynite grains indicate small amounts of $^{20}$Ne$_{tr}$, with the highest amounts observed in Mask1. Consequently, even though the amounts are too low to resolve the trapped component, the maskelynites show trapped Ne which is not present in the bulk material. Since maskelynite forms by shock-induced melting and quenching of plagioclase (Chen and El Goresy 2000; Tomioka and Miyahara 2017), it is possible that parent body noble gases were trapped during maskelynite formation and, due to rapid cooling of the melt, preserved as gas vesicles in the glassy material. Hence, it seems worthwhile to also analyze the maskelynite grains for the heavier noble gases Ar-Xe in order to provide evidence and potentially allow characterization of trapped noble gases in the parent body of NWA 11042.

Concentrations of Ar-Xe are generally small in bulk NWA 11042, as this achondrite has been intensely degassed, which is rather typical for thermally processed achondrites. The elemental abundances of trapped Ar and Kr relative to Xe ($^{36}$Ar/$^{132}$Xe = 702 and 680, $^{84}$Kr/$^{132}$Xe = 2.97 and 6.19, for NWA 11042 L and S, respectively) represent an Ar-rich or subsolar phase (Crabb and Anders 1982; Busemann and Eugster 2002) mixed with minor amounts of Q and air.

Kr isotopic ratios indicate a significant relative amount of Kr$_{cos}$ mixed with Kr$_{tr}$ (consistent with Q and air; solar wind is excluded based on its lack of light solar noble gases). Both samples also show additional excesses of $^{80}$Kr (pronounced) and $^{82}$Kr (less well resolved, as the excess/trapped Kr ratio is much lower for $^{82}$Kr than for $^{80}$Kr). For the determination of $^{83}$Kr$_{cos}$, we assumed $^{86}$Kr to be entirely trapped, ($^{80-83}$Kr/$^{86}$Kr)$_{tr}$ to cover both Q and air, and applied a typical ($^{80/82}$Kr/$^{83}$Kr)$_{cos}$ ratio for L-chondrites (Lavielle and Marti 1988) to determine ($^{80}$Kr/$^{82}$Kr)$_{excess}$ ratios. The ratio is 2.9 ± 1.8 for NWA 11042 L and, hence, in good agreement with the published value of 2.6 (Marti et al. 1966) that would result from (n, γ) reactions predominantly caused by neutrons in the energy interval of 30-300 eV. This shows that the excesses of $^{80}$Kr and $^{82}$Kr are likely neutron-induced from $^{79}$Br and $^{81}$Br, respectively. Since these comparably slow neutron energies require greater depths, the occurrence of neutron-induced isotopes implies a sufficiently large size of the meteoroid with a radius most likely >30 cm (Eberhardt et al. 1963).

The ratios of the lighter Xe isotopes also suggest considerable amounts of Xe$_{cos}$ relative to Xe$_{tr}$ (Q and air). Furthermore, we observe a $^{129}$I-derived excess of $^{129}$Xe. The concentration of $^{129}$Xe is remarkably similar in both samples although $^{129}$I is usually heterogeneously distributed in only a few carrier minerals in small samples. The same holds for $^{40}$Ar$_{rad}$ for which the concentration is also constant within NWA 11042 L and S. This could indicate the presence of large pockets of melt during the formation of NWA 11042, accompanied by degassing of the originally perhaps heterogeneously distributed carriers of $^{129}$I and homogenization of Xe (and $^{40}$Ar$_{rad}$).

### 3.6.2 Production Rates and Cosmic Ray Exposure Ages

We determined production rates (Table 6) of cosmogenic $^{3}$He, $^{21}$Ne, and $^{38}$Ar for the bulk samples based on the cosmogenic shielding parameter given here as $^{21}$Ne/$^{22}$Ne, marked with * in Table 3, and the bulk chemical composition of NWA 11042 determined in this work. The applied model (Leya and Masarik 2009) allows a large range of possible meteoroid sizes and sample depths. We considered pre-atmospheric meteoroid radii from 20 to 150 cm (a radius of 150 cm corresponds already to a meteoroid of ~45 t). Larger meteoroid sizes with >150 cm radius can likely be excluded: Only a few shielding depths (one per given radius 200, 300 and 500 cm, respectively, randomly distributed in large depths between 180 and 350 cm) would be possible, while for radii between 20 and 150 cm, the shielding conditions are in a consistent, restricted and -as expected- shallow range between 4 and 9 cm right below the surface of the meteoroid. The randomly distributed unrelated matches at large depths of a large meteoroid are very likely artifacts of less well-determined parameters in the model predictions for large shielding, low particle flux and very small cosmogenic production.

The preferred range of CRE ages for the two bulk samples deduced from cosmogenic $^{3}$He, $^{21}$Ne, and $^{38}$Ar concentrations are given in Table 7. The ranges are essentially identical, which suggests no preferential loss of $^{3}$He e.g. during close orbital approaches to the Sun. The overall preferred range of CRE ages for NWA 11042 is 30-46 Ma which does not include the uncertainty of the physical model predictions assumed to be ~15-20 %. Comparisons with CRE ages from other L chondrites will work best if the same production rate systematics are used. The less-precisely determined CRE age deduced from $^{83}$Kr$_{cos}$ and $^{126}$Xe$_{cos}$ based on the production rate given by Eugster (1988) are ~44 and ~55 Ma, respectively. Hence, they tend towards higher ages than the average value obtained from the lighter cosmogenic noble gases. This trend could indicate loss during meteoroid transfer in space affecting especially the lighter noble gases. However, the very consistent ages received from cosmogenic He, Ne, and Ar rather argue against such an effect. We used the error-weighted mean of the concentrations in NWA 11042 L and S for both $^{83}Kr_{cos}$ and $^{126}Xe_{cos}$.

The same method as for the bulk samples has been used to determine the production of cosmogenic He and Ne nuclides in the maskelynite samples. However, this turned out to be less straightforward. The model predicted $(^{21}Ne/^{22}Ne)_{cos}$ ratios of about 0.67 (standard deviation = 0.05) based on the measured maskelynite chemistry (Table 1). This is not consistent with the ratios measured in the maskelynite samples (between 0.81 and 0.85 for Mask1-3). Hence, it was not possible to determine the production rates and cosmic ray exposure ages for the maskelynite.

The model prediction of very low $(^{21}Ne/^{22}Ne)_{cos}$ ratios can be expected for maskelynite with its high Na content (here ~7 wt%). Such low ratios have already been measured in previous studies, e.g., for martian maskelynites (Schwenzer et al. 2007). Hence, the discrepancy between predicted and measured ratios suggests a deviation of the actual chemistry of Mask1-3 from pure maskelynite chemistry as given in Table 1. Reasons for this deviation could be an insufficient separation of the maskelynites used for noble gas analysis. When we use a mixture of maskelynite with more dominant minerals of bulk chemistry (60-85 wt%) in the model, we actually obtain a number of possible $(^{21}Ne/^{22}Ne)_{cos}$ ratios that are consistent with the measured ratios. Images of the maskelynite-rich separates taken before noble gas analysis further strengthen this assumption: the visible contamination of the maskelynite grains with bulk material seems to increase in the order of Mask2 > Mask3 > Mask1, which is proportional to the measured $(^{21}Ne/^{22}Ne)_{cos}$ ratios.

### 3.6.3 Radiogenic Gas Retention Ages

The nominal gas retention ages for both bulk samples (Table 7) derived from radiogenic $^4$He and bulk U and Th are in good agreement (T4, Table 3). The same holds for the retention ages received from $^{40}$Ar and bulk K (T40, Table 4). The T4 ages (average = 157 Ma) are generally much younger than the T40 ages (average upper limit for the K-Ar system 678 Ma and more reliably, ~484 Ma using the Ar-Ar data, see above) indicating further smaller heating event(s) after the reset of the $^{40}$K-$^{40}$Ar chronology system. The T4 ages for the maskelynites vary between 89-648 Ma which may partly be ascribed to the application of the average concentration of U and Th for maskelynite in NWA 11042. The ages obtained from both K-Ar (assuming no $^{40}$Ar air contamination) and Ar-Ar dating, are, at face value, not consistent. If we assume no air contamination (our preferred option based on the very similar $^{40}$Ar in both samples, see above) in order to achieve a similar $^{40}$K-$^{40}$Ar age as determined with the much more reliable $^{40}$Ar-$^{39}$Ar method, about 870 ppm of K is required to produce the measured $^{40}$Ar$_{rad}$ concentration, instead of the 581 ppm of K (Table 1) used here.

## 4.0 Discussion

The oxygen isotope composition of NWA 11042 suggests that it was derived from the L chondrite parent body. The bulk REE pattern (Fig. 7) is roughly 2×CI chondritic and slightly elevated above that of ordinary chondrites (Nakamura 1974). The compositions of the major phases, such as the Fo content of the olivine, the En content of the pyroxene, and the An content of the plagioclase, are also within the range of the L chondrites. The Cr isotopic composition of NWA 11042 also plots within the field of the ordinary chondrites (Fig. 2). These lines of evidence would put it in the category of the L-melt meteorites such as Chico and Patuxent Range (PAT) 91501, which crystallized from impact melted L chondrite material (Bogard et al. 1995; Mittlefehldt and Lindstrom 2001). However, other lines of evidence obscure such a succinct classification. NWA 11042 does not contain any relict chondrules within any of the areas studied. L-melt rocks sometimes contain multiple lithologies, including unmelted chondritic material of metamorphic grade 3-6 along with melted areas (Bischoff et al. 2006). Since NWA 11042 is composed of a single igneous lithology, as evident from the thin section and hand sample, an impact-derived origin must have been energetic enough to initiate complete melting in the region this meteorite was formed. The compositions of the major phases and textures of the unshocked portions of NWA 11042 resemble PAT 91501, which is also highly depleted in metal relative to the ordinary chondrites. PAT 91501 was suggested to originate from a deep impact-melt sheet or dike on the L chondrite parent body (Mittlefehldt and Lindstrom 2001; Benedix et al. 2008).

The trace element composition of NWA 11042 is similar to other examples of impact melted L-chondrites. Both NWA 11042 and the L impact melt PAT 91501 are depleted in siderophile elements (Fig. 9). These include highly siderophile elements (HSE) such as Ir, Pt, and Au, which fractionate readily during partial melting and extraction of even a low-degree of Fe-Ni-S liquid, but also moderately siderophile elements such as P, Co, Ni, Ga, and Ge. In particular, Ga/Al is depleted by a factor of ~2, and Ge/Si by a factor of ~100 relative to ordinary chondrites. These elements are more difficult to fractionate than HSE and their depletions are suggestive of moderate levels of partial melting enough to completely segregate the metal and silicate. However, the flat REE pattern of NWA 11042 suggests that the silicate portion experienced minimal loss of silicate melt relative to ordinary chondrites. Interestingly, the bulk REE pattern of PAT 91501 shows a slight negative Eu anomaly, implying that it lost some plagioclase by partial silicate melting.

Some volatile lithophile elements are also depleted in NWA 11042 relative to ordinary chondrites (Fig. 12). The K/Rb, K/La, and K/Th depletions indicate lower volatile contents than ordinary chondrites. However, NWA 11042 is enriched relative to the ordinary chondrites in Tl, which is an even more volatile element than K (Lodders 2003). While we cannot rule out that this might be due to weathering, it argues against simple devolatilization of the L chondrite parent body by impact melting.

The whole rock Ar-Ar age of 478.5 ± 5.5 Ma and the maskelynite Ar-Ar age of 484.0 ± 1.5 Ma fall within the range of many other Ar-Ar ages of shocked L chondritic material, which show a broad distribution peak between ~400 and 600 Ma (Bogard 2011; Swindle et al. 2014). These data suggest a large breakup event of the L chondrite parent body around this time, whose age has been refined to 470 ± 6 Ma by the separation of radiogenic Ar from various excess Ar components in several shocked L chondrites (Korochantseva et al. 2007). Additional measurements of the shocked L6 chondrite NWA 091 yielded an age of 475.4 ± 5.9 Ma after removal of excess Ar components (Weirich et al. 2012). Both of these studies used Cd shielding during irradiation of samples in order to resolve different components. The meteorite Ar-Ar ages can be linked to the terrestrial stratigraphic record, which shows a spike in fossil meteoritic material bearing L chondritic signatures at 467.3 ± 1.6 Ma (Korochantseva et al. 2007; Heck et al. 2008; Schmitz 2013). Lindskog et al. (2017) refined the age of the breakup event to 468.0 ± 0.3 Ma using Pb-Pb dating of detrital zircons in a limestone bed with an age of 467.50 ± 0.28 Ma and the CRE ages of the fossil meteoritic chromites found in the bed. Chromites found in these beds all have CRE ages under ~1.2 Ma (Heck et al. 2008; Schmitz 2013), and they are found to be progressively older proceeding up the stratigraphic section. The oxygen isotopic composition of >99% of these grains is L-chondritic (Heck et al. 2016). This suggests that they originated from a single impact event and transited rapidly from the asteroid belt to Earth-crossing orbits.

The Ar-Ar age of the maskelynite in NWA 11042 of 484.0 ± 1.5 Ma contributes an enigmatic data point to the narrative of L chondrite breakup event, previously established by Ar-Ar dating. While the stratigraphic record gives a precise age for a breakup event that delivered L chondritic material to Earth, the various Ar-Ar ages in the literature span a wide diversity of ages and uncertainties even within single lithologies (Fig. 13). The works of Korochantseva et al. (2007) and Weirich et al. (2012) were major breakthroughs in pinning down a single age for the breakup event, but they notably relied on corrections for trapped Ar to achieve plateau ages. By contrast, the Ar-Ar age of NWA 11042 contains minimal non-radiogenic Ar and thus does not require a trapped correction. Its plateau also converges on an age with a smaller error than any previously measured L chondritic material to our knowledge (Fig. 9). The plateau represents 92% of measured Ar, and as such it does not show any evidence for partial resetting or contamination. Interestingly, the only meteorite of the five presented by Korochantseva et al. (2007) that had a clear plateau without requiring a trapped Ar correction was Mbale, whose age of 479 ± 7 Ma is within error of our measurement for NWA 11042.

The high level of shock in NWA 11042, evident from the presence of ringwoodite in its melt pockets, is suggestive of a large impact event such as the breakup of the L chondrite parent body. The shock event and its associated heating must have been transient enough to preserve ringwoodite within the shock melt pockets, suggesting that the Ar-Ar age is representative of the event itself and not a later cooling age. Therefore, it is possible that the L chondrite parent body experienced a catastrophic impact event at 484.0 ± 1.5 Ma, and a large fragment was later disrupted causing an influx of material to Earth 15 Ma later. Only the second event seems to be present in the terrestrial fossil record. A fragment containing NWA 11042 would have then been disrupted much later, around 33 Ma ago, as evident from its exposure age, eventually bringing this meteorite to Earth. However, it is also possible that the older age is due to analytical difficulties involving the $^{40}$K decay constant, as discussed below. For the older age to be reflective of a second disruption event, the fragment containing NWA 11042 would have had to move away from the resonance that likely allowed the other fragment, involved in the 468 Ma disruption, to delivery material to Earth. After this, NWA 11042 would have had to move back towards such a resonance after the 33 Ma event that dislodged it from the fragment. Such a scenario is fairly complicated and must be weighed with other sources of inconsistency.

Although the analytical uncertainty of the maskelynite is reported at ±1.5 Ma, there are other important sources of uncertainty that require consideration when comparing variable geochronology data sets. For $^{40}Ar/^{39}Ar$ dating, error in the age of dating standard Fish Canyon tuff sanidine and the $^{40}K$ decay constant, as well as laboratory calibration issues, can be important. Prominent work by Kuiper et al. (2008) and Renne et al. (2010, 2011) yield variable estimates of Fish Canyon age and decay constants as well as their associated uncertainties. However, because the age of the standard is balanced by the decay constant, there is no significant change in age between the two calibrations. For instance, recalculating our reported age of 484.0 Ma based on Kuiper et al. (2008) to those of Renne et al. (2011) yields an age of 484.7 Ma. The total decay constant error reported by Renne et al. (2011) is substantially smaller than the error reported by Min et al. (2000) that is embodied in the Kuiper et al. (2008) age of Fish Canyon sanidine. Approximately ±0.1 Ma additional error is imparted by the Renne et al. (2011) estimate of decay constant uncertainty, whereas this would be ~±1 Ma for the Min et al. (2000) value. Finally, the NM Tech lab has not yet adopted the values of Renne et al. (2011) due to calibration issues between the lab and the Berkeley Geochronology Center. Various published and unpublished datasets indicate about 0.1 Ma offset at the K-Pg boundary that likely stems from lab procedural differences (e.g. Niespolo et al. 2017). If the offset is linearly extrapolated to 484 Ma, lab calibration errors would be about ±0.5 to 1 Ma.

In summary, it is difficult to accurately quantify calibration and constant errors, but conservatively our analytical uncertainty at ±1.5 Ma could expand to about 3-4 Ma given the magnitude of the numbers reported above. Although it is intriguing to consider two separable impact events on the L chondrite parent, given multiple sources of uncertainty we cannot definitively conclude this at this time.

The Sm-Nd age of NWA 11042 is less precise than the Ar-Ar age and reveals an ancient crystallization age completely different from the shock age (Fig. 10). The shock-melt pockets that plot off the isochron have most likely been partially reset by the shock event that the Ar-Ar systematics dates. Unfortunately, these partially reset components likely contaminated the rest of the mineral separates as well, as it would have been impossible to completely remove them during separation. The whole rock portion, meanwhile, obviously contains these melted components. The age of 4100 ± 160 Ma is thus only a minimum crystallization age for NWA 11042, and the true age must be older.

The thermal history of the L chondrites and their parent body, revealed by Ar-Ar chronometry of meteorites, records a large number of impacts between 400 and 600 Ma, with few between then and 4300 Ma, and a large number of ages around 4400-4500 Ma (Swindle et al. 2014). If the Sm-Nd age of NWA 11042 is taken to be "older than 4100 ± 160 Ma," then it is possible that its formation resulted from a large impact on the L chondrite parent body during this early time in solar system history. While the equilibrium temperature of 1111 ± 38 °C recorded by the pyroxenes in NWA 11042 is higher than the modelled maximum temperatures of impact-heated ordinary chondrite parent bodies by ~200 °C (Ciesla et al. 2013), its geochemical and petrological similarity to PAT 91501, determined to be an impact melt (Benedix et al. 2008), suggest that it could also be formed by impact melting.

It has been observed that as the H chondrites increase in metamorphic grade, their radiogenic ages determined by Hf-W, phosphate $^{207}$Pb-$^{206}$Pb, and phosphate fission track dating get increasingly younger (Trieloff et al. 2003; Kleine et al. 2008; Blackburn et al. 2017). A similar trend in phosphate $^{207}$Pb-$^{206}$Pb ages has been observed in the L chondrites as well (Blackburn et al. 2017). The "onion-shell" model of ordinary chondrite parent bodies (Trieloff et al. 2003) suggests that the cause of this progressive decrease in age is that the U and Pb-bearing material in more highly metamorphosed chondrites took longer to reach closure temperature because they were deeper within their parent body and thus experienced a longer heating period. An alternative hypothesis to an impact melting origin is that NWA 11042, and potentially the ungrouped achondrite NWA 4284, could represent samples of material located even deeper within the "onion" which was partially melted and differentiated. Partially melted material from this event has to this point not been identified in the meteorite collection, although interestingly the oxygen isotopes of silicate inclusions in the IVA iron meteorites also plot in the L chondrite field (Clayton and Mayeda 1996), and their chromium isotopes are similar to those of ordinary chondrites (Sanborn et al. 2018). Other lines of evidence for primary melting of chondrite parent bodies include modelling results that suggest the possibility of planetesimal melting in the presence of insulating chondritic crusts (Elkins-Tanton et al. 2011; Weiss and Elkins-Tanton 2013; Fu and Elkins-Tanton 2014), chronological evidence that suggests planetesimal melting and differentiation was concurrent with chondrule formation (Amelin 2008; Kleine et al. 2009; Schiller et al. 2011; Connelly et al. 2012; Srinivasan et al. 2018), and remanent magnetization identified in ordinary chondrites suggesting they formed in the presence of a core dynamo (Shah et al. 2017).

However, an undisturbed onion-shell model for ordinary chondrite parent bodies is inconsistent with varying metallographic cooling rates measured from diffusion profiles across FeNi metal grains in H and L chondrites (Taylor et al. 1987; Scott et al. 2014), which do not show a monotonic relationship with metamorphic grade. Instead, early-formed parent bodies were likely disrupted by impacts prior to cooling below the temperature at which FeNi diffusion becomes negligible (~500 °C). Importantly, this does not negate the radiochronometric evidence for an onion shell; it instead suggests that originally thermally stratified bodies were disrupted and reaccreted into "rubble piles" which preserved the radioisotopic imprint but reset the metallographic cooling history. If NWA 11042 represents deep partial melting of an onion-shell parent body, then it must have melted before subsequent disruption.

Wu and Hsu (2019) recently dated NWA 11042 using phosphate $^{207}$Pb-$^{206}$Pb and found disturbed concordance with an upper intercept age of 4,479 ± 43 Ma and a lower intercept age of 465 ± 47 Ma. These ages concord with phosphate $^{207}$Pb-$^{206}$Pb ages obtained from the L6 chondrite Novato of 4,472 ± 31 Ma and 473 ± 38 Ma (Yin et al. 2014). The younger ages, interpreted as shock resetting ages, agree well with our more precise Ar-Ar age of 484.0 ± 1.5 Ma. The older age of Wu and Hsu (2019), interpreted as the crystallization age, argues strongly against petrogenesis by partial differentiation of the L chondrite parent body, since any short-lived radiogenic heat sources would have decayed away by that time. Instead, Wu and Hsu (2019) suggest that NWA 11042 was formed inside a thick melt dike within a large impact structure which remained thermally insulated for long enough to generate the igneous properties observed in the meteorite.

The petrologic characteristics of NWA 11042 and NWA 4284 (this work) suggest that their parent body experienced a low degree of partial melting, enough to extract an Fe-Ni-S liquid but not enough to mobilize low-degree silicate melts that would have preferentially extracted plagioclase. This is consistent with the sub-liquidus temperature recorded by the pyroxene thermometer. Plagioclase extraction would have also created a bulk composition depleted in light REE with a negative Eu anomaly. Additionally, the Fe/Mn in the olivine and pyroxene in NWA 11042 and 4284, along with the Fo# in the two rocks and ordinary chondrites is virtually identical. Were the two achondrites melt residues produced by fractional melting, a trend in Fe/Mn with olivine Mg# would be expected (Sunshine et al. 2007). Instead, a low degree of partial melt and extraction of an Fe-Ni-S liquid are likely responsible for the depletion of siderophile elements and the slight change in modal mineralogy observed between the two meteorites.

Overall, the petrological, geochemical, and chronological evidence presented here and by Wu and Hsu (2019) suggest that NWA 11042 was originally formed by a deep impact melt vein or dike that imparted its igneous characteristics upon the L chondrite parent body. Incidentally, while the asteroids 8 Flora and 6 Hebe are associated with the LL and H chondrites, respectively (Gaffey and Gilbert 1998; Vernazza et al. 2008), there is currently no large asteroid linked to the L chondrites. It is therefore likely that this parent body was disrupted, and the shock ages of L chondritic material and terrestrial meteorite records attest to this. While NWA 11042 likely represents an impact melt rock sourced from the L chondrite parent body, the possibility of primary igneous melting on ordinary chondrite parent bodies remains a possibility that cannot be ruled out based on current knowledge and available samples.

## Conclusions

We have performed petrologic, geochemical, isotopic, and chronological analyses on NWA 11042, a cumulate rock with affinities to L chondrites. These data were compared to the petrology of NWA 4284 and the geochemistry of PAT 91501. Oxygen and Cr isotopes, along with bulk petrology and geochemistry suggest that NWA 11042 formed from the partial melting of the L chondrite parent body at least $4.10 \pm 0.16$ Ga ago, but likely earlier. The $^{207}Pb$-$^{206}Pb$ crystallization age of $4,479 \pm 43$ Ma obtained by Wu and Hsu (2019) is likely more correct. This partial melting event was likely caused by an impact large enough to form a deep insulating melt vein or dike that allowed slow cooling and the formation of a cumulate texture. The removal of an Fe-Ni-S liquid caused heavy depletion of siderophile elements but left no evidence of fractional melting in the lithophile elements, similar to the L impact melt PAT 91501. The Ar-Ar chronometer was reset at $484.0 \pm 1.5$ Ma, implicating this meteorite with the breakup of the L chondrite parent body which is recorded in a spike of Ar-Ar ages of L chondritic material between 400 and 600 Ma and the terrestrial fossil record at ~468 Ma. While the precision of this Ar-Ar age and that of the fossil chromite record could suggest the possibility of multiple large impacts during the breakup, uncertainties within various geochronology parameters prevent a definitive conclusion of multiple events from being reached.

Acknowledgements

This article was made possible in part through funding provided by a NASA grant, Research Opportunities in Space and Earth Sciences. Additional funding was received through the New Mexico Space Grant Consortium. We acknowledge the UNM Electron Microbeam Facility and Dr. Michael Spilde, as well as the UNM Center for Microanalysis and Dr. Adrian Brearley for their invaluable assistance in gathering electron microanalytical data. We also acknowledge the UNM X-Ray Diffraction Lab and Dr. Eric Peterson for providing XRD analyses. This work was in part supported by the Swiss SNF including the NCCR "Planet S" (D. K. & H. B.).

Figures

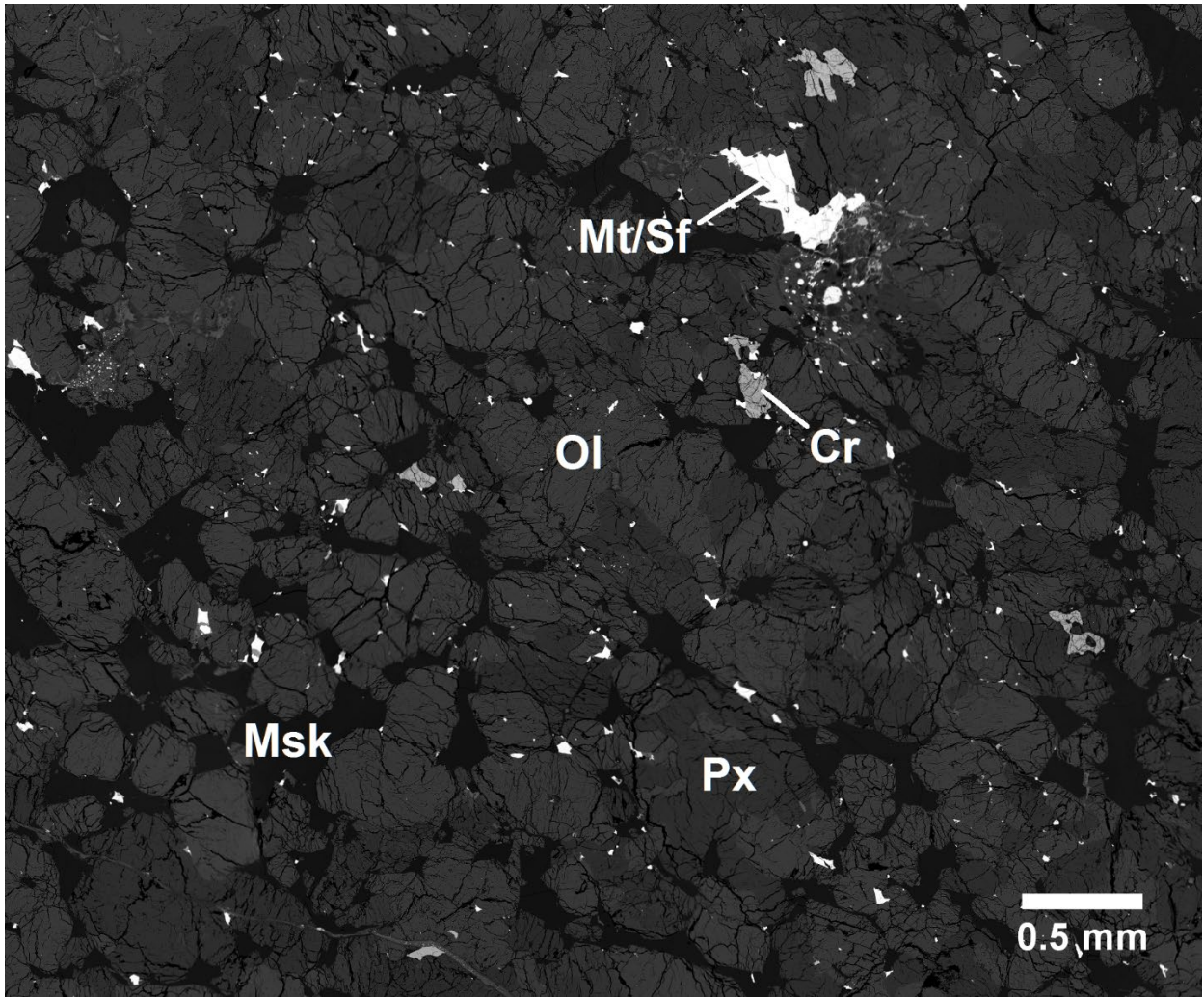

**Fig. 1.** BSE image of NWA 11042. Ol=olivine, Px=pyroxene, Msk=maskelynite, Cr=chromite, Mt/Sf=FeNi metal/sulfide.

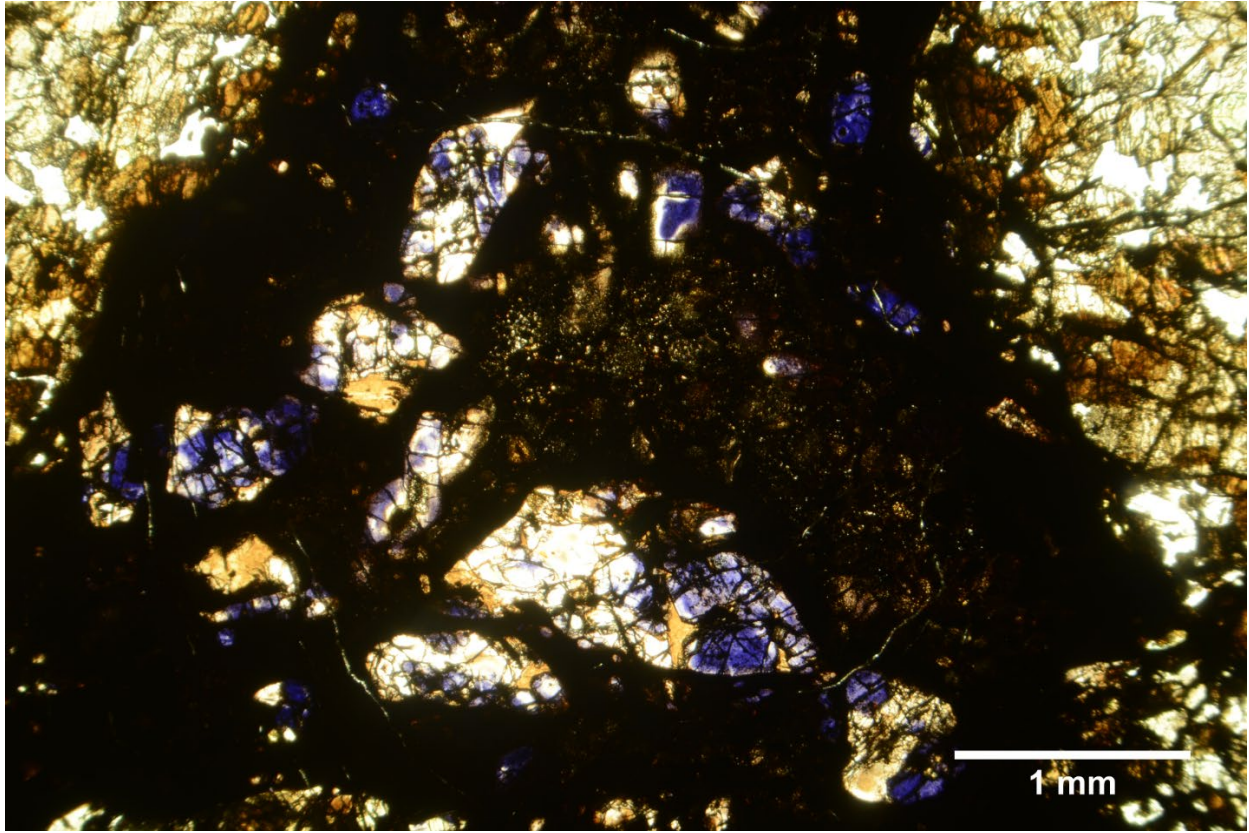

**Fig. 2.** Plane-polarized light (PPL) image of large shock-melt pocket in NWA 11042. It contains partially melted phases and some of the olivine has been transformed to blue ringwoodite.

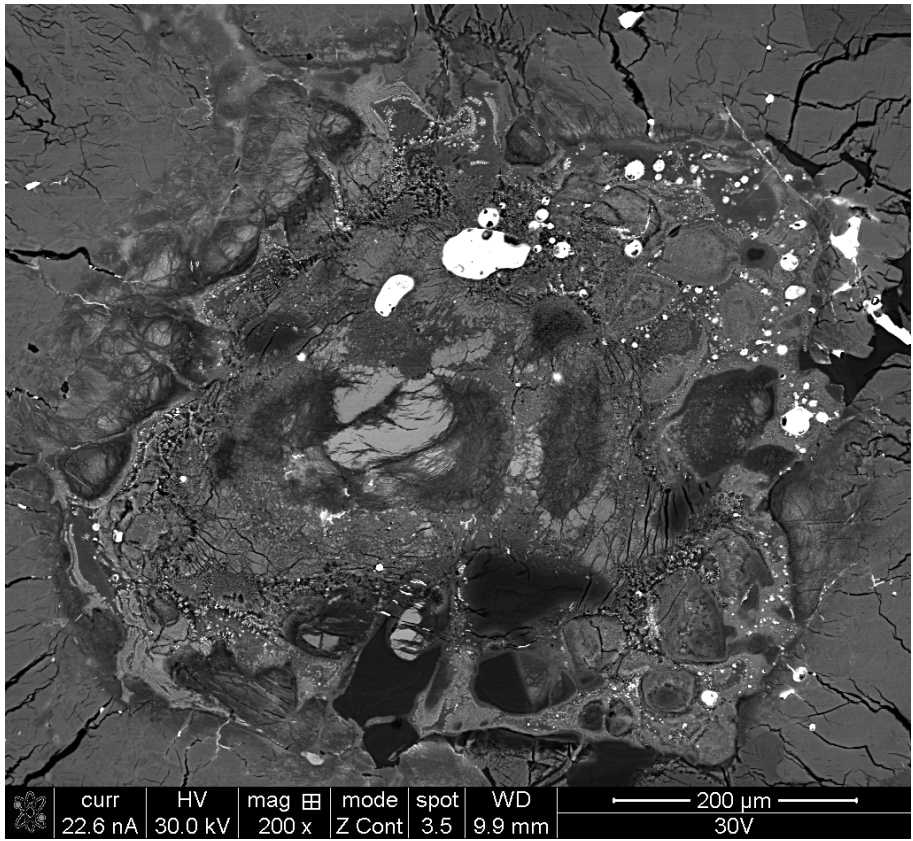

**Fig. 3.** BSE image of shock-melt pocket in NWA 11042. A dark reduction rim surrounds the pocket and exsolved metal nanoparticles are observed in adjacent mineral grains.

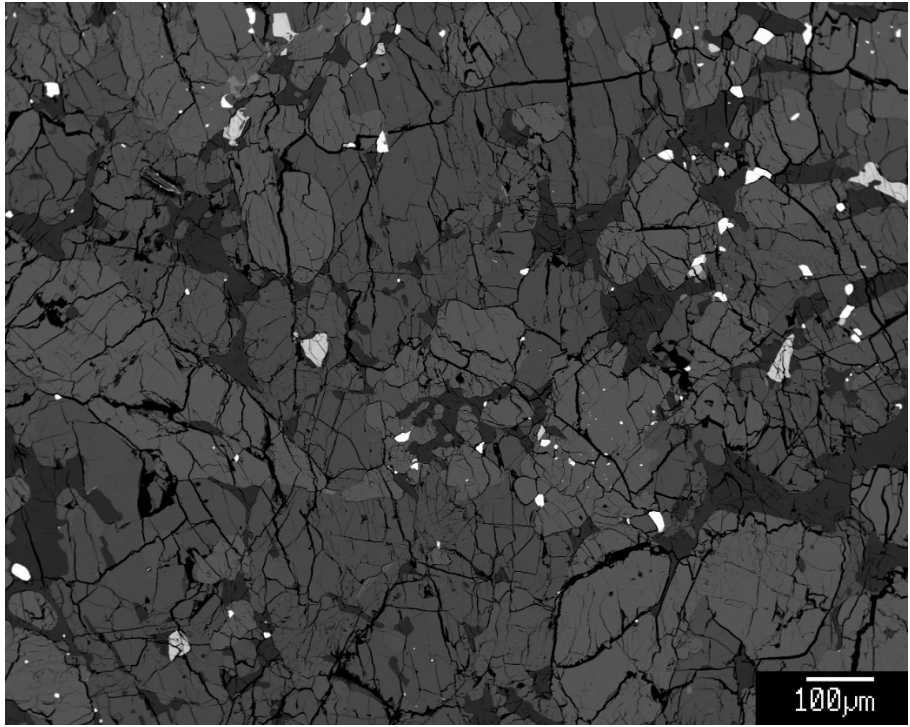

**Fig. 4.** BSE image of NWA 4284. Phases are, in ascending order of brightness, plagioclase, pyroxene, olivine, chromite, metal/sulfide. Some of the plagioclase (center of image) appears vitrified.

|  | Olivine | Opx | Cpx | Mask. | Chromite | Melt | Bulk |
|---|---|---|---|---|---|---|---|
| n | 5 | 5 | 3 | 5 | 1 | 6 | 4 |
| $SiO_2$ | 36.85 | 53.49 | 51.30 | 62.19 | 0.15 | 38.31 | 46.52 |
| $TiO_2$ | 0.01 | 0.15 | 0.30 | 0.07 | 3.04 | 0.05 | 0.11 |
| $Al_2O_3$ | 0.05 | 0.78 | 1.22 | 20.31 | 9.17 | 2.13 | 2.39 |
| $FeO_T$ | 23.46 | 14.45 | 8.29 | 2.36 | 31.50 | 24.74 | 17.86 |
| MnO | 0.49 | 0.46 | 0.32 | 0.04 | 0.71 | 0.32 | 0.43 |
| MgO | 37.99 | 27.14 | 18.78 | 2.19 | 3.34 | 26.38 | 28.60 |
| CaO | 0.38 | 2.18 | 17.39 | 3.04 | n.d. | 1.89 | 2.26 |
| $Na_2O$ | 0.55 | 0.42 | 0.67 | 8.31 | 0.57 | 0.81 | 1.02 |
| $K_2O$ | 0.08 | 0.05 | 0.03 | 0.77 | 0.18 | 0.23 | 0.07 |
| $P_2O_5$ | 0.02 | 0.03 | 0.02 | 0.12 | 0.00 | 0.04 | 0.08 |
| S | 0.08 | 0.17 | 0.54 | 0.58 | 0.09 | 4.93 | 0.29 |
| $Cr_2O_3$ | 0.05 | 0.69 | 1.13 | 0.01 | 51.33 | 0.17 | 0.35 |
| Total | 100.00 | 100.00 | 100.00 | 100.00 | 100.07 | 100.00 | 100.00 |
| Mg# | 74.3 | 77.0 | 80.2 | N/A | 15.9 | 64.1 | 74.1 |
| FeO/MnO | 48.3 | 31.3 | 25.6 | N/A | 44.6 | 61.5 | 41.7 |
| K/Rb | 468 | 329 | 240 | 225 | 148 | 321 | 207 |
| Be | 0.027 | 0.041 | 0.040 | 0.057 | 0.013 | 0.028 | 0.031 |
| B | 1.32 | 1.03 | 1.10 | 1.28 | 1.21 | 4.40 | 0.62 |
| V | 20 | 133 | 320 | 2 | 3620 | 40 | 74 |
| Co | 24 | 15 | 18 | 6.4 | 18 | 45 | 25 |
| Ni | 54 | 21 | 128 | 36 | 33 | 313 | 92 |
| Cu | 4.8 | 2.7 | 5.5 | 2.6 | 5.8 | 121 | 4.9 |
| Zn | 51 | 39 | 15 | 3.3 | 3000 | 67 | 45 |
| Ga | 2.49 | 2.58 | 3.05 | 11.38 | 99.5 | 11.6 | 2.57 |
| Ge | 0.067 | 0.017 | 0.087 | 0.108 | n.d. | 0.309 | 0.122 |
| As | 0.407 | 0.047 | 0.060 | 0.043 | 0.060 | 0.839 | 0.058 |
| Rb | 1.40 | 1.18 | 1.04 | 28.5 | 10.2 | 5.35 | 3.01 |
| Sr | 6.97 | 2.25 | 10.1 | 111 | 1.03 | 65.3 | 13.0 |
| Y | 0.34 | 2.34 | 11.0 | 0.506 | n.d. | 0.83 | 2.82 |
| Zr | 0.71 | 3.21 | 8.85 | 11.1 | 5.32 | 1.40 | 6.34 |
| Nb | 0.087 | 0.285 | 0.413 | 0.480 | 0.835 | 0.316 | 0.482 |
| Cs | 0.178 | 0.193 | 0.115 | 0.367 | 0.791 | 0.259 | n.d. |
| Ba | 55.6 | 9.66 | 15.5 | 40.3 | 6.65 | 178 | 16.4 |
| La | 0.022 | 0.117 | 0.544 | 0.851 | 0.011 | 0.080 | 0.412 |
| Ce | 0.095 | 0.316 | 2.460 | 1.587 | 0.023 | 0.196 | 1.262 |
| Pr | 0.018 | 0.042 | 0.485 | 0.178 | n.d. | 0.027 | 0.160 |
| Nd | 0.110 | 0.183 | 2.902 | 0.680 | 0.310 | 0.159 | 0.820 |
| Sm | 0.038 | 0.093 | 1.253 | 0.138 | n.d. | 0.054 | 0.274 |
| Eu | 0.010 | 0.010 | 0.091 | 0.831 | 0.044 | 0.097 | 0.104 |
| Gd | 0.051 | 0.209 | 1.843 | 0.142 | n.d. | 0.089 | 0.397 |
| **Tb** | 0.012 | 0.052 | 0.354 | 0.021 | 0.003 | 0.018 | 0.074 |
| **Dy** | 0.079 | 0.396 | 2.361 | 0.122 | 0.019 | 0.136 | 0.523 |
| **Ho** | 0.018 | 0.096 | 0.482 | 0.025 | 0.009 | 0.037 | 0.117 |
| **Er** | 0.088 | 0.337 | 1.393 | 0.068 | 0.055 | 0.117 | 0.359 |
| **Tm** | 0.016 | 0.059 | 0.184 | 0.012 | 0.023 | 0.024 | 0.059 |
| **Yb** | 0.128 | 0.382 | 1.087 | 0.055 | n.d. | 0.169 | 0.358 |
| **Lu** | 0.024 | 0.065 | 0.155 | 0.009 | n.d. | 0.036 | 0.063 |
| **Hf** | 0.025 | 0.087 | 0.297 | 0.324 | 0.138 | 0.040 | 0.216 |
| **Ta** | 0.003 | 0.023 | 0.027 | 0.034 | 0.043 | 0.011 | 0.030 |
| **W** | 0.005 | 0.005 | 0.018 | 0.011 | 0.010 | 0.012 | 0.061 |
| **Re** | n.d. | 0.000 | 0.000 | n.d. | 0.012 | 0.001 | 0.006 |
| **Os** | 0.017 | 0.001 | 0.001 | 0.001 | n.d. | 0.006 | 0.006 |
| **Ir** | 0.009 | 0.001 | 0.001 | 0.002 | 0.001 | 0.007 | 0.006 |
| **Pt** | 0.018 | n.d. | 0.001 | 0.000 | 0.009 | 0.012 | 0.005 |
| **Au** | 0.002 | 0.001 | 0.002 | 0.002 | n.d. | 0.005 | 0.025 |
| **Tl** | 0.007 | 0.002 | 0.005 | 0.001 | 0.009 | 0.178 | 0.013 |
| **Pb** | 0.044 | 0.036 | 0.039 | 0.033 | 0.080 | 0.092 | 0.092 |
| **Bi** | 0.000 | 0.001 | 0.000 | 0.002 | n.d. | 0.001 | 0.010 |
| **Th** | 0.004 | 0.039 | 0.048 | 0.083 | n.d. | 0.016 | 0.063 |
| **U** | 0.040 | 0.019 | 0.028 | 0.034 | 0.018 | 0.215 | 0.038 |
| **Nb/Ta** | 28.77 | 12.44 | 15.31 | 14.16 | 19.22 | 29.71 | 16.05 |
| **Zr/Hf** | 28.03 | 36.85 | 29.81 | 34.17 | 38.51 | 34.68 | 29.30 |
| **Th/U** | 0.09 | 2.07 | 1.70 | 2.41 | n.d. | 0.08 | 1.64 |
| **Ni/Ge** | 809 | 1215 | 1473 | 332 | n.d. | 1011 | 748 |
| **Ga/Al** | 0.01037 | 0.00063 | 0.00047 | 0.00011 | 0.00205 | 0.00103 | 0.00020 |
| **Ge/Si** | 3.90E-07 | 6.91E-08 | 3.64E-07 | 3.72E-07 | n.d. | 1.73E-06 | 5.63E-07 |

**Table 1.** Major oxide and trace elemental abundances for NWA 11042 measured with LA-ICP-MS. Oxides and S are in wt%; trace elements are in ppm. Laser ablation spot sizes are 50-100 μm for individual phases, and bulk measurements consist of averages of rastered patterns with a 150 μm spot size. n.d. = not detected.

|  | Olivine | Opx | Cpx | Plag. | Bulk |
|---|---|---|---|---|---|
| n | 5 | 5 | 3 | 2 | 8 |
| $SiO_2$ | 39.95 | 52.37 | 52.67 | 63.18 | 43.16 |
| $TiO_2$ | 0.07 | 0.03 | 0.26 | 0.26 | 0.12 |
| $Al_2O_3$ | 1.17 | 0.29 | 1.11 | 20.23 | 1.98 |
| $FeO_T$ | 23.59 | 15.89 | 8.59 | 1.61 | 19.96 |
| MnO | 0.44 | 0.42 | 0.34 | 0.03 | 0.43 |
| MgO | 32.62 | 28.81 | 18.30 | 1.43 | 29.70 |
| CaO | 0.35 | 1.25 | 16.94 | 2.40 | 2.30 |
| $Na_2O$ | 0.63 | 0.07 | 0.42 | 9.21 | 1.03 |
| $K_2O$ | 0.10 | 0.02 | n.d. | 1.00 | 0.10 |
| $P_2O_5$ | 0.14 | 0.02 | 0.02 | 0.32 | 0.21 |
| S | 0.77 | 0.05 | 0.00 | 0.29 | 0.44 |
| $Cr_2O_3$ | 0.17 | 0.78 | 1.34 | 0.04 | 0.57 |
| Total | 100.00 | 100.00 | 100.00 | 100.00 | 100.00 |
| Mg# | 58.0 | 64.6 | 68.1 | N/A | 59.8 |
| FeO/MnO | 54.0 | 37.3 | 25.6 | N/A | 46.8 |
| K/Rb | 255 | 133 | 167 | 286 | 296 |
| Be | n.d. | n.d. | n.d. | 0.215 | n.d. |
| B | 0.60 | n.d. | n.d. | 2.36 | 0.38 |
| V | 32 | 99 | 306 | 8 | 88 |
| Co | 63 | 35 | 16 | 38 | 53 |
| Ni | 528 | 108 | 22 | 308 | 300 |
| Cu | 20 | 2.4 | 0.6 | 8.9 | 11 |
| Zn | 65 | 49 | 22 | 4.0 | 49 |
| Ga | 2.259 | 0.909 | 4.085 | 31.21 | 3.782 |
| Ge | 0.351 | 0.080 | 0.076 | 0.427 | 0.211 |
| As | 0.036 | 0.026 | 0.010 | n.d. | 0.043 |
| Se | 2.444 | 0.142 | n.d. | 1.586 | 1.256 |
| Rb | 3.395 | 0.674 | n.d. | 29.20 | 2.863 |
| Sr | 3.756 | n.d. | 5.715 | 89.30 | 9.471 |
| Y | 1.340 | 0.538 | 8.962 | 2.200 | 2.763 |
| Zr | 9.116 | 0.377 | 2.672 | 29.44 | 6.426 |
| Nb | 0.504 | 0.075 | 0.023 | 2.506 | 0.440 |
| Mo | 0.044 | n.d. | n.d. | 0.329 | 0.054 |
| Ru | n.d. | n.d. | n.d. | n.d. | 0.006 |
| Rh | 0.002 | n.d. | n.d. | 0.019 | 0.001 |
| Pd | 0.003 | n.d. | n.d. | 0.024 | 0.010 |
| Ag | 0.013 | 0.004 | n.d. | 0.022 | 0.024 |
| Cd | 0.103 | 0.134 | 0.111 | 0.149 | 0.015 |
| Sn | 0.039 | 0.005 | 0.016 | 0.015 | 0.034 |
| Sb | 0.001 | n.d. | 0.005 | 0.003 | 0.009 |
| Te | 0.152 | 0.030 | n.d. | 0.188 | 0.031 |
| Cs | 0.346 | n.d. | n.d. | 1.457 | 0.067 |
| Ba | 1.482 | n.d. | n.d. | 28.14 | 3.381 |
| La | 0.203 | 0.006 | 0.173 | 0.568 | 0.387 |
| Ce | 0.479 | 0.028 | 0.609 | 1.156 | 1.126 |
| **Pr** | 0.077 | 0.005 | 0.165 | 0.177 | 0.157 |
| **Nd** | 0.444 | 0.041 | 1.144 | 1.004 | 0.814 |
| **Sm** | 0.111 | 0.019 | 0.714 | 0.338 | 0.268 |
| **Eu** | 0.033 | 0.006 | 0.038 | 0.642 | 0.062 |
| **Gd** | 0.159 | 0.036 | 1.240 | 0.378 | 0.402 |
| **Tb** | 0.030 | 0.009 | 0.240 | 0.067 | 0.071 |
| **Dy** | 0.201 | 0.086 | 1.683 | 0.455 | 0.491 |
| **Ho** | 0.050 | 0.028 | 0.363 | 0.103 | 0.109 |
| **Er** | 0.168 | 0.078 | 1.078 | 0.300 | 0.324 |
| **Tm** | 0.025 | 0.017 | 0.161 | 0.042 | 0.048 |
| **Yb** | 0.189 | 0.119 | 0.956 | 0.275 | 0.316 |
| **Lu** | 0.044 | 0.021 | 0.163 | 0.036 | 0.052 |
| **Hf** | 0.209 | 0.009 | 0.158 | 0.945 | 0.195 |
| **Ta** | 0.023 | 0.001 | n.d. | 0.149 | 0.024 |
| **W** | 0.093 | 0.010 | 0.011 | 0.240 | 0.095 |
| **Re** | n.d. | n.d. | n.d. | 0.004 | 0.002 |
| **Os** | 0.001 | 0.000 | n.d. | 0.004 | 0.005 |
| **Ir** | 0.003 | 0.002 | 0.001 | n.d. | 0.006 |
| **Pt** | 0.007 | 0.002 | n.d. | n.d. | 0.006 |
| **Au** | n.d. | n.d. | n.d. | 0.024 | 0.003 |
| **Tl** | 0.000 | 0.000 | 0.001 | 0.001 | 0.001 |
| **Pb** | 0.029 | 0.005 | 0.003 | 0.089 | 0.050 |
| **Bi** | 0.001 | n.d. | n.d. | n.d. | 0.002 |
| **Th** | 0.044 | 0.004 | 0.003 | 0.170 | 0.052 |
| **U** | 0.015 | 0.002 | 0.001 | 0.050 | 0.017 |
| **Nb/Ta** | 14.13 | 11.72 | 5.95 | 16.41 | 18.74 |
| **Zr/Hf** | 18.37 | 26.89 | 17.16 | 30.67 | 32.96 |
| **Th/U** | n.d. | 3.07 | n.d. | 3.97 | 3.15 |
| **Ni/Ge** | n.d. | 1433 | n.d. | 841 | 1422 |
| **Ga/Al** | n.d. | 0.00064 | 0.00069 | 0.00029 | 0.00036 |
| **Ge/Si** | 1.86E-06 | 3.32E-07 | 3.08E-07 | 1.44E-06 | 1.05E-06 |

**Table 2**. Major oxide and trace elemental abundances for PAT 91501 measured with LA-ICP-MS using the same methods as for NWA 11042.

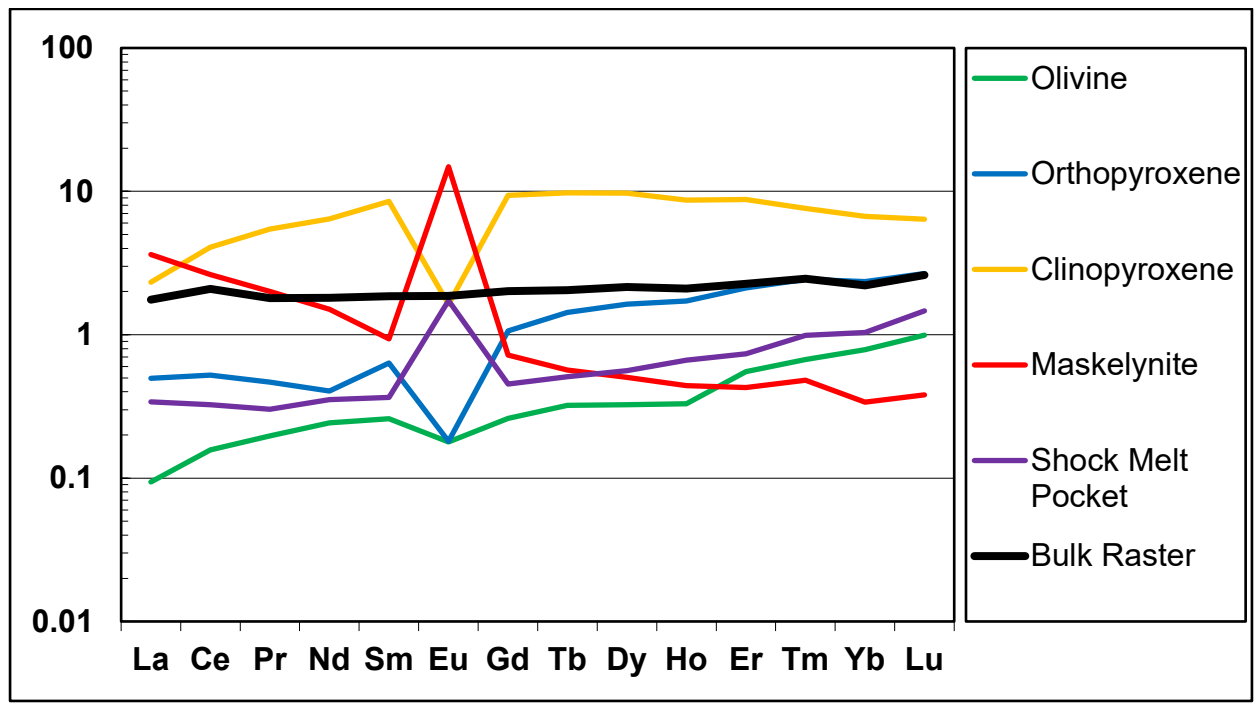

**Fig. 5** showing REE concentrations for various phases in NWA 11042 along with a bulk rock raster, normalized to CI chondritic abundances (Anders and Grevesse 1989).

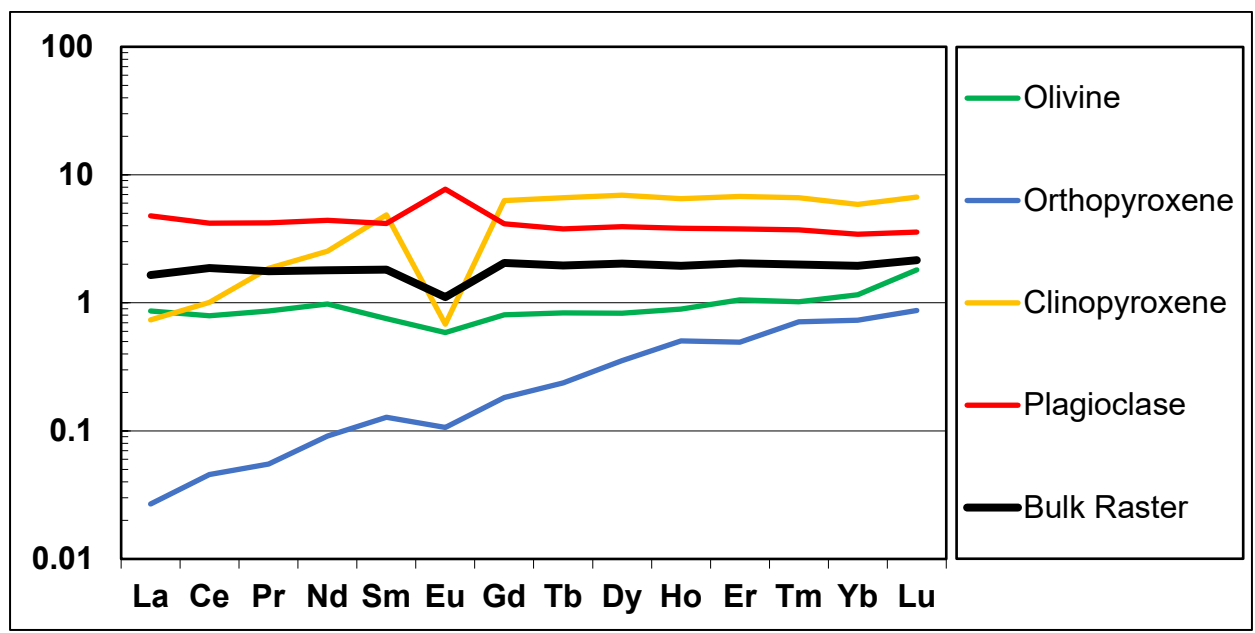

**Fig. 6** showing REE concentrations for various phases in PAT 91501 along with a bulk rock raster, normalized to CI chondritic abundances (Anders and Grevesse 1989).

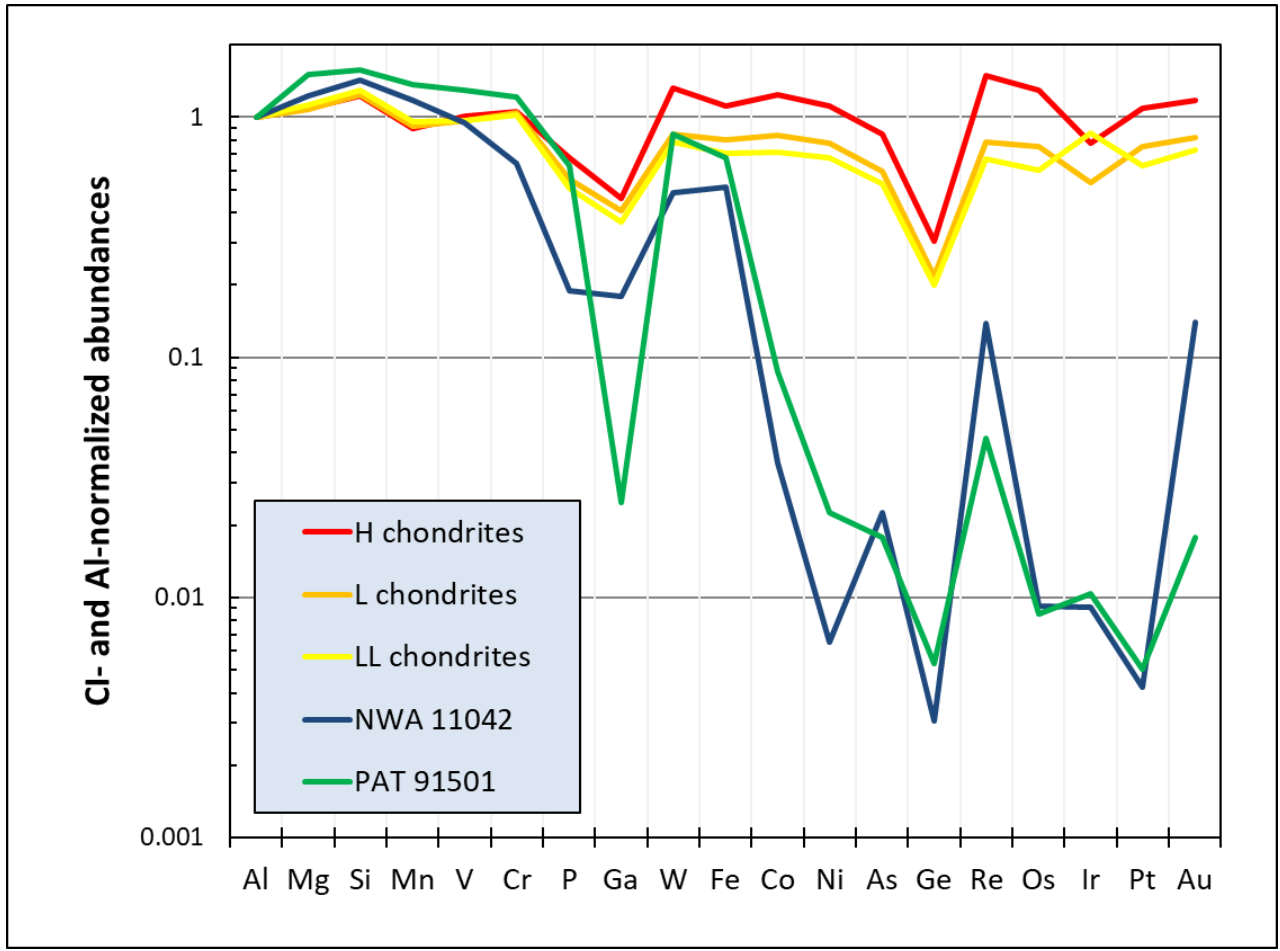

**Fig. 7** Elemental abundances plotted in order of increasing siderophile character from left to right, for NWA 11042, PAT 91501, and the ordinary chondrites (Wasson and Kallemeyn 1988). Abundances are doubly normalized to the refractory element Al and to CI chondrites (Anders and Grevesse 1989).

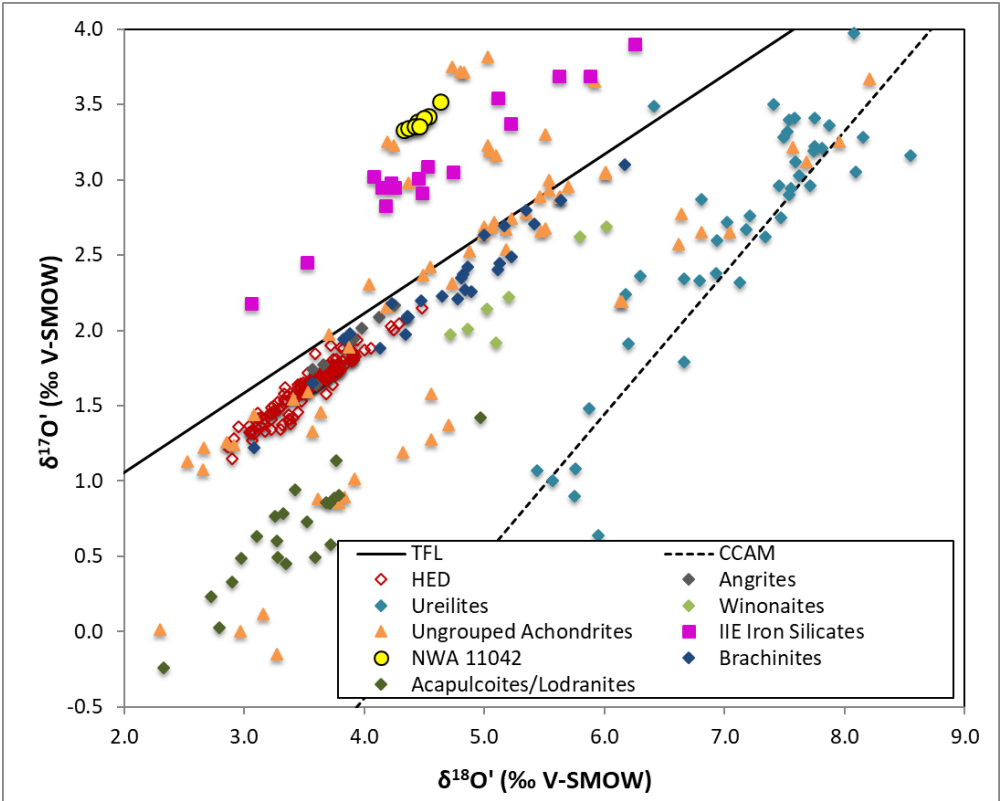
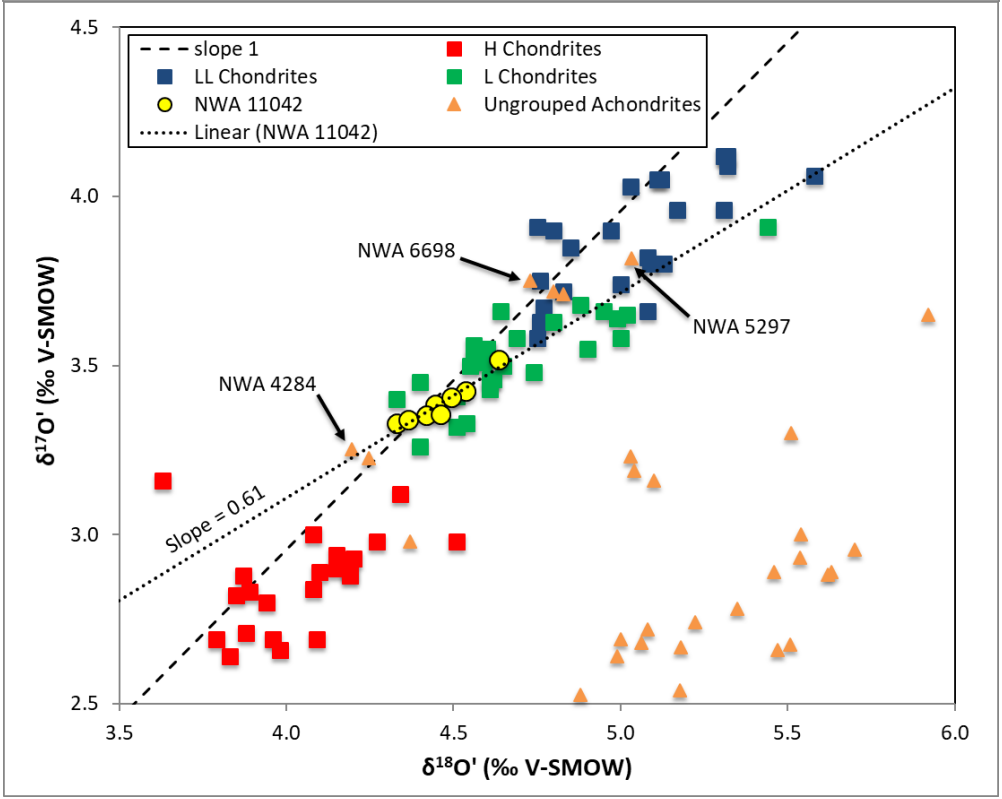

**Fig. 8.** (Top) Triple oxygen plot for grouped and ungrouped achondrites and primitive achondrite NWA 11042. NWA 11042 plots within the domain of the ordinary chondrites. (Bottom) Truncated view of triple oxygen plot showing NWA 11042, ordinary chondrites, and ungrouped achondrites. A linear regression through the data for NWA 11042 is also shown, along with the slope=1 line. TFL = terrestrial fractionation line (slope=0.528), CCAM = carbonaceous chondrite anhydrous mineral line. Oxygen isotopic data is from the MetBull Meteoritic database and Clayton et al. (1984, 1991); Clayton and Mayeda (1996, 1999); Greenwood et al. (2005, 2012, 2017).

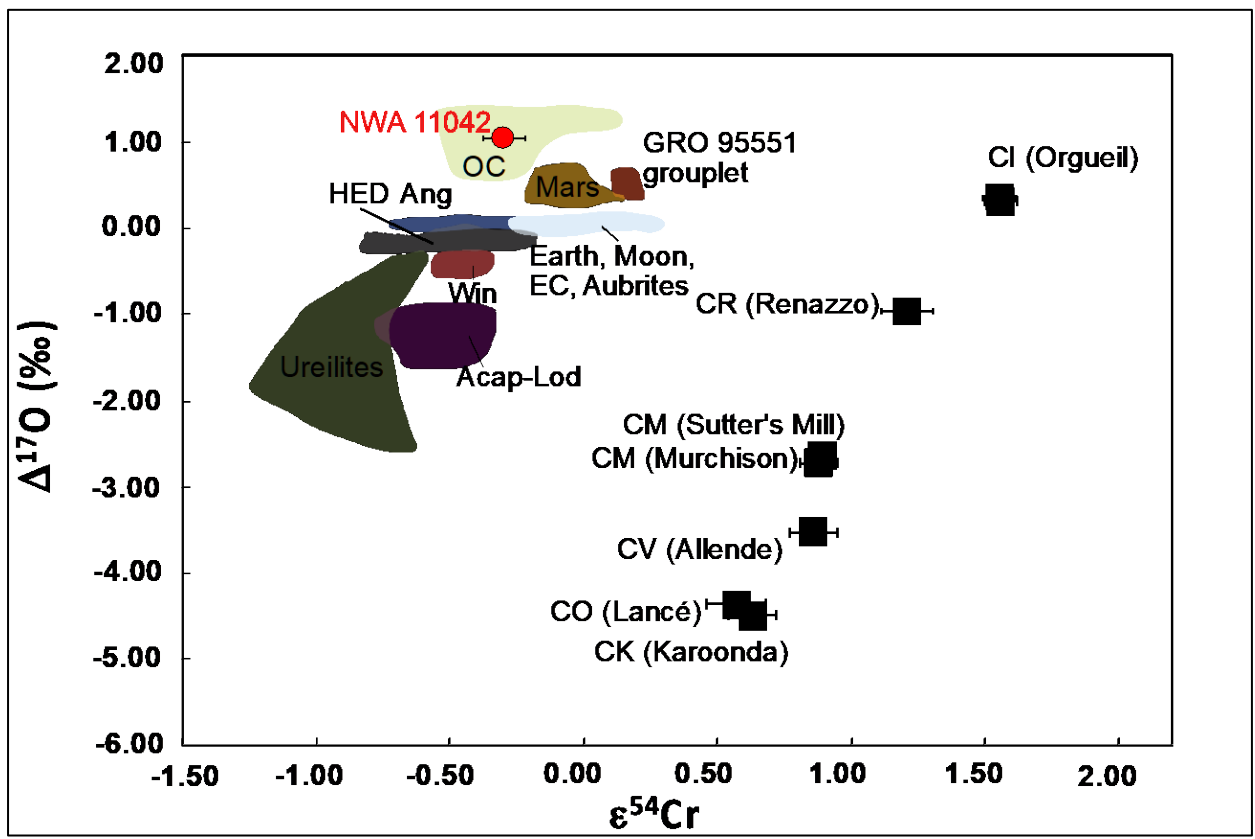

**Fig. 9.** $\varepsilon^{54}Cr$ versus $\Delta^{17}O$ diagram plotting NWA 11042 and other meteorite groups (carbonaceous and non-carbonaceous). Shaded regions indicated compositional ranges for particular non-carbonaceous meteorite groups. Literature data for $\Delta^{17}O$ values are from Clayton et al. 1984, 1991, Clayton and Mayeda 1996, 1999; Greenwood et al. 2005; Jenniskens et al. 2012, 2014; Popova et al. 2013. Data for $\varepsilon^{54}Cr$ are from Shukolyukov and Lugmair 2006; Ueda et al. 2006; Trinquier et al. 2007; Yin et al. 2009; Jenniskens et al. 2012, 2014; Popova et al. 2013.

| Sample | Type | Mass [mg] | $^4$He | $^3$He/$^4$He [$10^{-4}$] | $^4$He$_{rad}$ | $^3$He$_{cos}$[a] | $^{20}$Ne | $^{20}$Ne/$^{22}$Ne | ($^{21}$Ne/$^{22}$Ne)$_{cos}$[a] | $^{20}$Ne$_{tr}$ | $^{21}$Ne$_{cos}$[a] |
|---|---|---|---|---|---|---|---|---|---|---|---|
| NWA 11042 L | Bulk | 100.380 ± 0.010 | 443 ± 4 | 1323 ± 24 | 112 ± 27 | 58.6 ± 0.9 | 11.6 ± 0.1 | 0.848 ± 0.003 | 0.908 ± 0.006 | n.d. | 12.44 ± 0.08 |
| NWA 11042 S | Bulk | 19.873 ± 0.014 | 437 ± 2 | 1402 ± 9 | 91 ± 28 | 61.2 ± 0.2 | 12.2 ± 0.1 | 0.840 ± 0.004 | 0.895 ± 0.003* | n.d. | 12.96 ± 0.07 |
| Mask1 | Maskelynite Separate | 0.1143 ± 0.0008 | 548 ± 5 | 341 ± 11 | 443 ± 10 | 18.7 ± 0.6 | 5.20 ± 0.06 | 1.302 ± 0.015 | 0.810 ± 0.011 | 2.20 ± 0.06 | 3.23 ± 0.04 |
| Mask2 | Maskelynite Separate | 0.0794 ± 0.0006 | 219 ± 2 | 1300 ± 42 | 58 ± 13 | 28.5 ± 0.9 | 7.36 ± 0.08 | 1.030 ± 0.010 | 0.850 ± 0.011 | 1.71 ± 0.32 | 6.07 ± 0.09 |
| Mask3 | Maskelynite Separate | 0.1021 ± 0.0008 | 447 ± 4 | 803 ± 26 | 244 ± 17 | 35.9 ± 1.2 | 8.71 ± 0.09 | 1.018 ± 0.009 | 0.847 ± 0.011 | 1.97 ± 0.10 | 7.25 ± 0.10 |

**Table 3** showing concentrations of He and Ne isotopes and resolved components as well as isotopic compositions. Concentrations in $10^{-8}$ cm$^3$ STP g$^{-1}$. Uncertainties on measured noble gas concentrations include ion counting statistics, uncertainties of sample mass and calibration gas amounts as well as variations in blank level and detector sensitivity. Uncertainties on isotopic ratios also include ion counting statistics and blank corrections, and instrumental mass discrimination. Uncertainties on cosmogenic and trapped concentrations include uncertainties related to the deconvolution when applied, i.e., the respective choice of cosmogenic and trapped endmember components and all experimental uncertainties. $^4$He$_{rad}$ uncertainties include uncertainties on $^3$He$_{cos}$ concentrations and the typical ($^3$He/$^4$He)$_{cos}$ ratio (Wieler 2002). N.d. = not detected.

[a] measured = cosmogenic concentration.
* Higher precision for ($^{21}$Ne/$^{22}$Ne)$_{cos}$ in NWA 11042 S compared to NWA 11042 L due to measurement with the electron multiplier instead of the Faraday cup. Ratio used as shielding indicator to determine the production rates for both bulk samples.

| Sample | $^{36}$Ar | $^{36}$Ar/$^{38}$Ar | $^{36}$Ar$_{tr}$ | $^{38}$Ar$_{cos}$ | $^{40}$Ar$_{rad}$[a] | $^{84}$Kr | $^{78}$Kr/$^{84}$Kr | $^{80}$Kr/$^{84}$Kr | $^{82}$Kr/$^{84}$Kr | $^{83}$Kr/$^{84}$Kr | $^{86}$Kr/$^{84}$Kr |
|---|---|---|---|---|---|---|---|---|---|---|---|
| NWA 11042 L | 1.56 ± 0.02 | 1.048 ± 0.004 | 0.76 ± 0.05 | 1.34 ± 0.01 | 183 ± 3 | 32 ± 1 | 0.0342 ± 0.0011 | 0.228 ± 0.007 | 0.341 ± 0.012 | 0.349 ± 0.013 | 0.279 ± 0.012 |
| NWA 11042 S | 1.36 ± 0.03 | 1.055 ± 0.025 | 0.67 ± 0.05 | 1.16 ± 0.01 | 186 ± 8 | 61 ± 7 | 0.0199 ± 0.0025 | 0.134 ± 0.016 | 0.291 ± 0.042 | 0.271 ± 0.039 | 0.289 ± 0.054 |

**Table 4** showing concentrations of $^{36}$Ar and $^{84}$Kr, Ar and Kr isotopic compositions and resolved components. Ar concentration in $10^{-8}$ cm$^3$ STP g$^{-1}$. Kr concentration in $10^{-12}$ cm$^3$ STP g$^{-1}$. See Table 3 for uncertainties.

[a] measured = radiogenic concentration.

| Sample | $^{132}$Xe | $^{124}$Xe/$^{132}$Xe | $^{126}$Xe/$^{132}$Xe | $^{128}$Xe/$^{132}$Xe | $^{129}$Xe/$^{132}$Xe | $^{130}$Xe/$^{132}$Xe | $^{131}$Xe/$^{132}$Xe | $^{134}$Xe/$^{132}$Xe | $^{136}$Xe/$^{132}$Xe |
|---|---|---|---|---|---|---|---|---|---|
| NWA 11042 L | 10.8 ± 0.4 | 0.0245 ± 0.0014 | 0.0380 ± 0.0016 | 0.131 ± 0.005 | 1.798 ± 0.068 | 0.169 ± 0.007 | 0.863 ± 0.043 | 0.425 ± 0.018 | 0.348 ± 0.017 |
| NWA 11042 S | 9.8 ± 2.6 | 0.0238 ± 0.0068 | 0.0347 ± 0.0096 | 0.130 ± 0.041 | 1.881 ± 0.565 | 0.174 ± 0.060 | 0.855 ± 0.317 | 0.457 ± 0.158 | 0.357 ± 0.125 |

**Table 5** showing concentration of $^{132}$Xe and Xe isotopic composition. Xe concentration in $10^{-12}$ cm$^3$ STP g$^{-1}$. See Table 3 for uncertainties.

| Type | P3 | P21 | P38 | Radius [cm] | Depth [cm] |
|---|---|---|---|---|---|
| Bulk | 1.345-1.907 | 0.285-0.392 | 0.032-0.044 | 20-150 | 4-9 |

**Table 6** showing possible shielding of the sample, i.e., meteoroid size and depth below surface and resulting "allowed" production rate ranges for cosmogenic $^3$He, $^{21}$Ne, and $^{38}$Ar. Production rates in $10^{-8}$ cm$^3$ STP g$^{-1}$ Ma$^{-1}$. Uncertainties on production rates contain the uncertainties of the noble gas measurements and shielding conditions as suggested by $(^{21}$Ne/$^{22}$Ne$)_{cos}$ = 0.895 ± 0.003. Uncertainties of the model are not included but believed to be 15-20%.

| Sample | T3 | T21 | T38 | T4 | T40 |
|---|---|---|---|---|---|
| NWA 11042 L | 30-44 | 31-44 | 31-42 | 173 ± 41 | 673 ± 9 |
| NWA 11042 S | 32-46 | 31-43 | 26-36 | 141 ± 43 | 682 ± 25 |
| **Total** | | 30-46* | | 157 ± 23 | 678 ± 6 |
| Mask1 | - | - | - | 648 ± 14 | - |
| Mask2 | - | - | - | 89 ± 20 | - |
| Mask3 | - | - | - | 365 ± 25 | - |
| Average | - | - | - | 367 ± 280 | - |

**Table 7** showing preferred ranges of cosmic ray exposure ages for cosmogenic $^3$He, $^{21}$Ne, and $^{38}$Ar, and gas retention ages for radiogenic $^4$He and $^{40}$Ar. All ages in Ma. Cosmic ray exposure age ranges include variations in cosmogenic concentrations (choice of endmembers) and mainly production rates (allowed shielding conditions determined by $(^{21}$Ne/$^{22}$Ne$)_{cos}$. Uncertainties on gas retention ages include uncertainties on radiogenic concentrations. Average values for T4 and T40 are given with standard deviation.

*Uncertainties of the model are not included but believed to be 15-20%.

|  | NWA 11042 | Ordinary Chondrites | | |
|---|---|---|---|---|
|  |  | H | L | LL |
| Fe (wt.%) | 13.9 | 28 | 22 | 19 |
| Fe/Si (atomic) | 0.32 | 0.81 | 0.57 | 0.52 |
| Metal (vol%) | 0.7 | 8.4 | 4.1 | 2 |
| Olivine Fa | 25 | 16-20 | 21-26 | 27-31 |
| Pyroxene Fs | 20 | 15-17 | 18-22 | 22-30 |
| $\Delta^{17}O$ | 1.03 | 0.7 | 1.1 | 1.3 |

**Table 8** Summary table of some properties of NWA 11042 compared to the ordinary chondrites. Ordinary chondrite data taken from Vernazza et al. (2015).

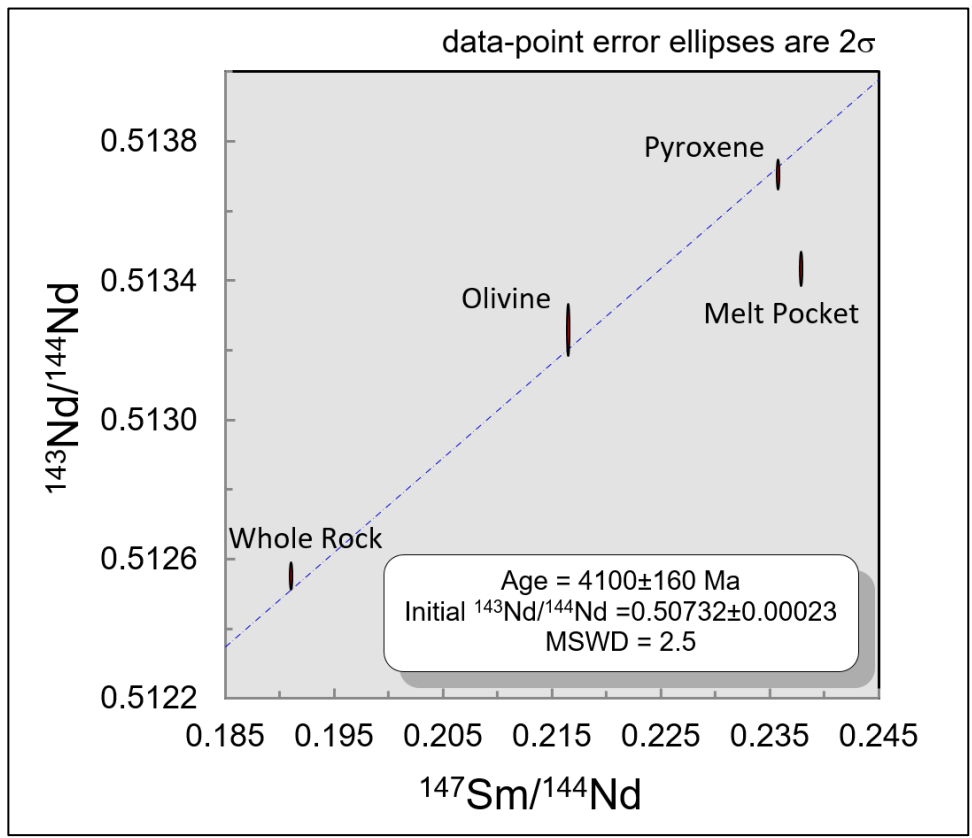

**Fig. 10** showing Sm-Nd isochron from mineral separates of NWA 11042. The melt pocket is not included in the calculation.

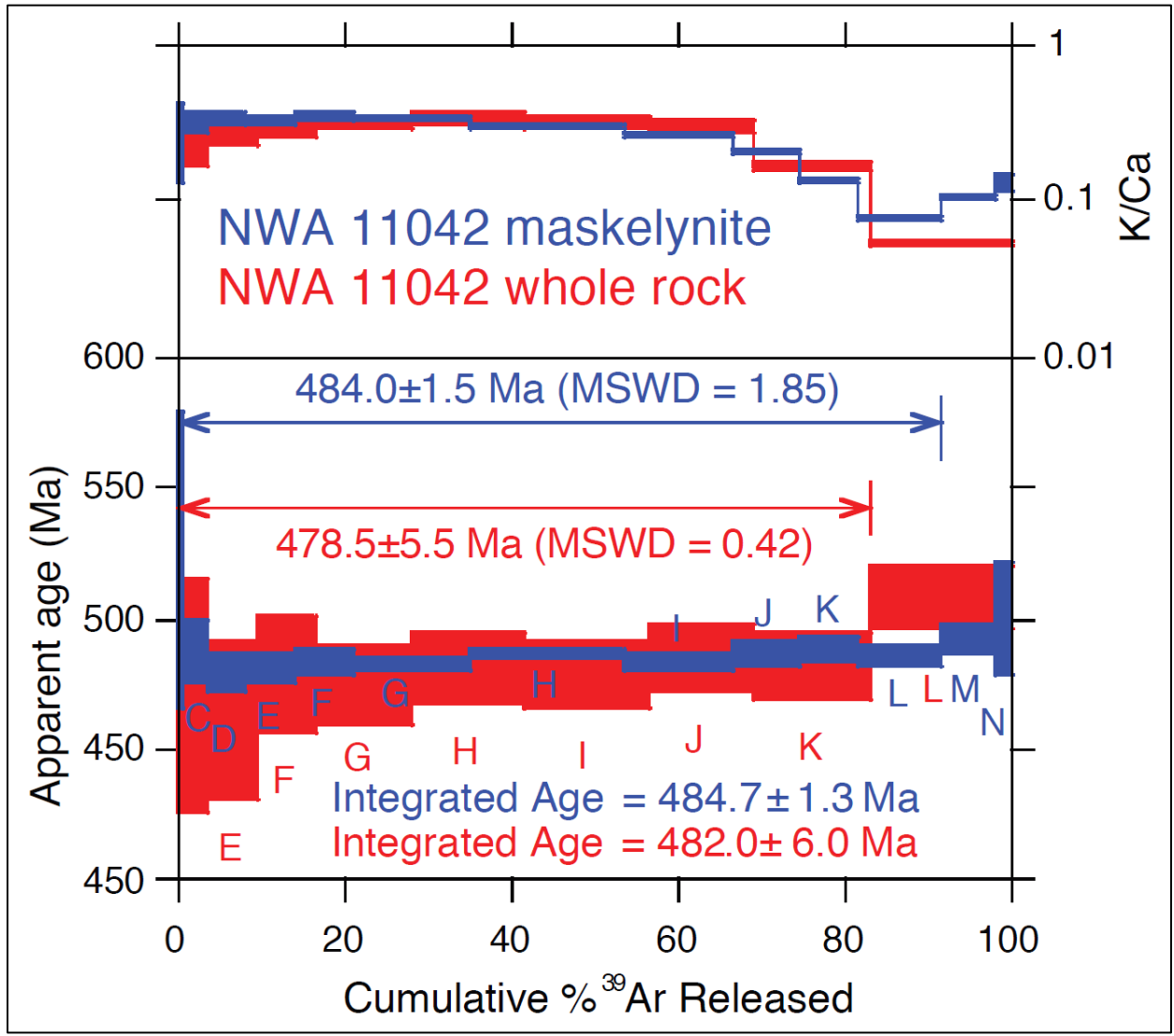

**Fig. 11** showing results of Ar-Ar dating of NWA 11042 (red=whole rock, blue=maskelynite). Uncertainties in ages are 2σ.

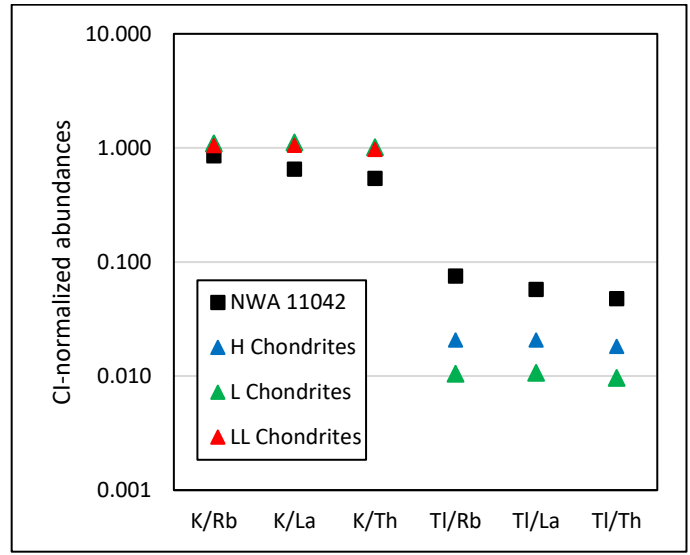

**Fig. 12** Volatile lithophile element to refractory lithophile element ratios for NWA 11042 and ordinary chondrites, normalized to CI chondrites (Wasson and Kallemeyn 1988; Anders and Grevesse 1989)

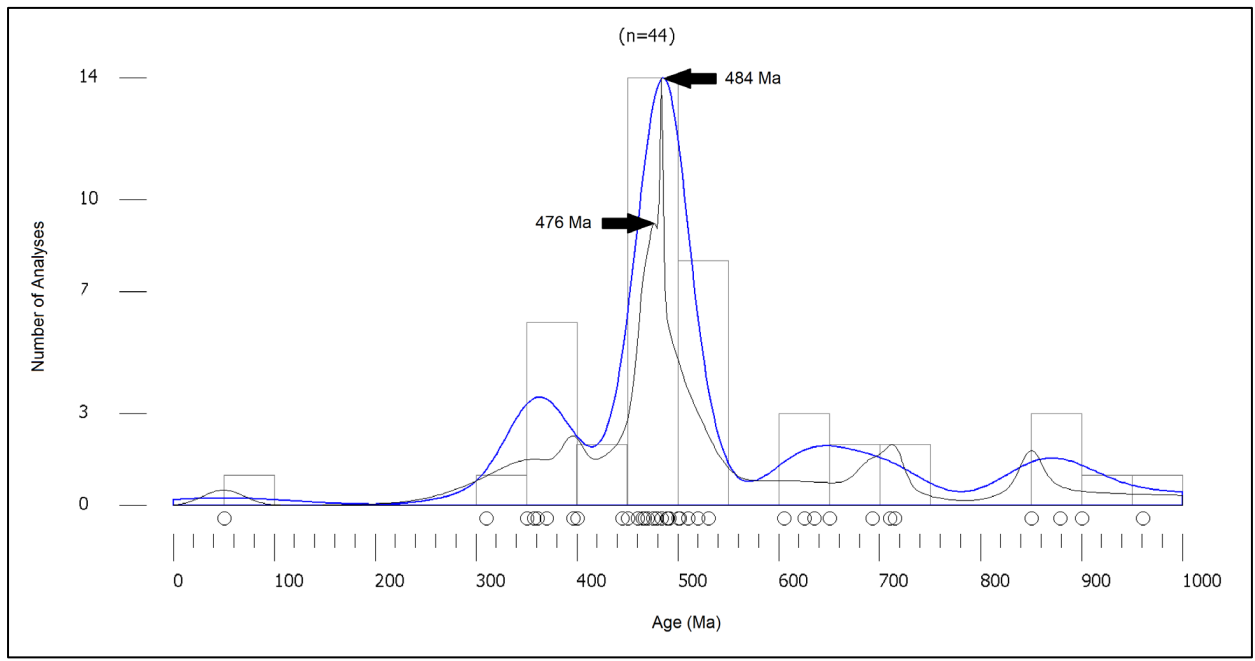

**Fig. 13**. Histogram of measured Ar-Ar ages for L chondritic material (grey bars), probability density function (PDF) (black line), and kernel density estimation (KDE) (blue line) (Vermeesch 2012) including the ages measured for NWA 11042. The PDF shows two peaks, the larger of which converges with the one broad peak of the KDE. Plot calculated using DensityPlotter (Vermeesch 2012) with a bin size of 50 Ma. Ar-Ar data are from Turner 1969; Bogard et al. 1976, 1995; Kaneoka et al. 1979, 1988; Bogard and Hirsch 1980; Vu Minh et al. 1984; McConville et al. 1988; Bogard 1995; Kring et al. 1996; Kunz et al. 1997; Sato et al. 2000; Swindle et al. 2006, 2011; Korochantseva et al. 2007; Metzler et al. 2011; Weirich et al. 2012; Ruzicka et al. 2015. Only plateau ages are included, and in their absence the age interpretations of Swindle et al. (2014) are used.

# Appendix

## Table A1 Ar-Ar ratios and age data

| | ID | Power (Watts) | $^{40}Ar/^{39}Ar$ | $^{38}Ar/^{39}Ar$ | $^{37}Ar/^{39}Ar$ | $^{36}Ar/^{39}Ar$ (x $10^{-3}$) | $^{39}Ar_K$ (x $10^{-15}$ mol) | K/Ca | $^{40}Ar^*$ (%) | $^{39}Ar$ (%) | Age (Ma) | ±1σ (Ma) |
|---|---|---|---|---|---|---|---|---|---|---|---|---|
| | \multicolumn{12}{l}{**NWA 11042, Whole Rock, 10.58 mg, J=0.0028597±0.02%, NM-290A, Lab#=65602-01**} |
| | D | 1.6 | 102.0 | 0.2606 | 2.472 | 250.4 | 0.2 | 0.21 | 100.0 | 3.9 | 469.1 | 22 |
| | E | 1.9 | 99.68 | 0.2783 | 1.868 | 249.2 | 0.3 | 0.27 | 100.0 | 9.6 | 459.6 | 15 |
| | F | 2.2 | 104.2 | 0.2625 | 1.742 | 231.2 | 0.3 | 0.29 | 100.0 | 16.9 | 478.1 | 11 |
| | G | 2.6 | 103.1 | 0.2257 | 1.648 | 196.6 | 0.5 | 0.31 | 100.0 | 28.1 | 473.6 | 7.7 |
| | H | 3.0 | 104.7 | 0.2350 | 1.506 | 184.2 | 0.6 | 0.34 | 100.0 | 41.9 | 480.0 | 6.5 |
| | I | 3.5 | 104.1 | 0.2503 | 1.628 | 200.6 | 0.7 | 0.31 | 100.0 | 56.7 | 477.4 | 6.1 |
| | J | 4.0 | 105.6 | 0.2733 | 1.743 | 241.3 | 0.6 | 0.29 | 100.0 | 69.5 | 483.5 | 6.4 |
| | K | 6.0 | 104.8 | 0.6063 | 3.140 | 603.1 | 0.7 | 0.16 | 100.0 | 83.3 | 480.6 | 6.4 |
| x | L | 10.0 | 110.8 | 1.715 | 9.513 | 1941.1 | 0.8 | 0.054 | 100.0 | 100.0 | 506.9 | 5.8 |
| | **Integrated age ± 2σ** | | | | n=9 | | 4.7 | 0.16 | | K2O=0.06% | 482.0 | 6.0 |
| | **Plateau ± 2σ** | steps D-K | | | n=8 | MSWD=0.42 | 3.9 | | | 83.3 | 478.5 | 5.5 |
| | \multicolumn{12}{l}{**NWA 11042, Maskelynite, 9.11 mg, J=0.0028819±0.02%, NM-290F, Lab#=65661-01**} |
| | B | 0.8 | 114.4 | 0.2471 | 1.876 | 211.3 | 0.2 | 0.27 | 100.0 | 0.7 | 522.1 | 28 |
| | C | 1.3 | 105.5 | 0.2195 | 1.622 | 215.7 | 0.8 | 0.31 | 100.0 | 3.8 | 486.5 | 5.9 |
| | D | 1.6 | 103.4 | 0.1823 | 1.560 | 163.4 | 1.2 | 0.33 | 100.0 | 8.4 | 478.0 | 3.8 |
| | E | 1.9 | 103.9 | 0.1765 | 1.563 | 148.6 | 1.6 | 0.33 | 100.0 | 14.4 | 479.8 | 2.9 |
| | F | 2.2 | 104.4 | 0.1825 | 1.483 | 142.1 | 1.8 | 0.34 | 100.0 | 21.3 | 481.9 | 2.5 |
| | G | 3.0 | 104.5 | 0.1952 | 1.560 | 151.0 | 3.7 | 0.33 | 100.0 | 35.2 | 482.2 | 1.2 |
| | H | 4.0 | 105.3 | 0.2090 | 1.718 | 157.3 | 4.9 | 0.30 | 100.0 | 53.5 | 485.78 | 0.92 |
| | I | 5.0 | 104.4 | 0.2279 | 1.934 | 170.0 | 3.4 | 0.26 | 100.0 | 66.5 | 482.2 | 1.4 |
| | J | 6.0 | 105.2 | 0.2854 | 2.513 | 221.4 | 2.1 | 0.20 | 100.0 | 74.5 | 485.7 | 2.1 |
| | K | 7.0 | 105.7 | 0.3719 | 3.832 | 280.1 | 1.9 | 0.13 | 100.0 | 81.8 | 487.9 | 2.4 |
| | L | 10.0 | 104.8 | 0.5462 | 6.522 | 403.1 | 2.6 | 0.078 | 100.0 | 91.7 | 484.9 | 1.8 |
| x | M | 15.0 | 106.4 | 0.4578 | 4.950 | 335.3 | 1.8 | 0.10 | 100.0 | 98.4 | 491.2 | 2.7 |
| x | N | 15.0 | 108.4 | 0.3772 | 4.029 | 254.9 | 0.4 | 0.13 | 100.0 | 100.0 | 499.0 | 10.6 |
| | **Integrated age ± 2σ** | | | | n=13 | | 26.5 | 0.19 | | $K_2O$=0.39% | 484.70 | 1.3 |
| | **Plateau ± 2σ** | steps B-L | | | n=11 | MSWD=1.85 | 26.1 | | | 91.8 | 484.0 | 1.5 |

**Notes:**
Isotopic ratios corrected for blank, radioactive decay, and mass discrimination, not corrected for interfering reactions.
Errors quoted for individual analyses include analytical error only, without interfering reaction or J uncertainties.
Integrated age calculated by summing isotopic measurements of all steps.
Integrated age error calculated by quadratically combining errors of isotopic measurements of all steps.
Plateau age is inverse-variance-weighted mean of selected steps.
Plateau age error is inverse-variance-weighted mean error (Taylor 1982) times root MSWD where MSWD>1.
Plateau error is weighted error of Taylor (1982).
Isotopic abundances after Steiger and Jäger (1977).
x preceding sample ID denotes analyses excluded from plateau age calculations.
Weight percent $K_2O$ calculated from $^{39}Ar$ signal, sample weight, and instrument sensitivity.
Ages calculated relative to FC-2 Fish Canyon Tuff sanidine interlaboratory standard at 28.201 Ma (Kuiper et al. 2008).
Decay Constant (LambdaK (total)) = 5.463e-10/a (Min et al. 2000).
Correction factors:
   $(^{39}Ar/^{37}Ar)_{Ca}$ = 0.00073 ± 0.000020
   $(^{36}Ar/^{37}Ar)_{Ca}$ = 0.0002725 ± 0.0000009
   $(^{38}Ar/^{39}Ar)_{K}$ = 0.012718±0.00008
   $(^{40}Ar/^{39}Ar)_{K}$ = 0.0088 ± 0.0004

**Table A2** Ar-Ar Intensity Data

| Run_Date | Material | $^{40}$Ar | $^{40}$Ar Er | $^{39}$Ar | $^{39}$Ar Er | $^{38}$Ar | $^{38}$Ar Er | $^{37}$Ar | $^{37}$Ar Er | $^{36}$Ar | $^{36}$Ar Er |
|---|---|---|---|---|---|---|---|---|---|---|---|
| 8/3/2017 | WR | 46.6640 | 0.0541 | 0.4576 | 0.0247 | 0.1192 | 0.0135 | 1.1314 | 0.0934 | 0.1146 | 0.0011 |
| 8/3/2017 | WR | 67.2400 | 0.0546 | 0.6745 | 0.0250 | 0.1878 | 0.0134 | 1.2598 | 0.1020 | 0.1681 | 0.0012 |
| 8/3/2017 | WR | 90.0965 | 0.0544 | 0.8642 | 0.0231 | 0.2269 | 0.0128 | 1.5054 | 0.0936 | 0.1998 | 0.0013 |
| 8/3/2017 | WR | 136.0018 | 0.0558 | 1.3187 | 0.0243 | 0.2976 | 0.0144 | 2.1734 | 0.0983 | 0.2592 | 0.0014 |
| 8/3/2017 | WR | 169.3004 | 0.0587 | 1.6163 | 0.0249 | 0.3799 | 0.0137 | 2.4338 | 0.0964 | 0.2978 | 0.0015 |
| 8/3/2017 | WR | 181.6001 | 0.0577 | 1.7447 | 0.0255 | 0.4367 | 0.0143 | 2.8399 | 0.0969 | 0.3501 | 0.0015 |
| 8/3/2017 | WR | 159.0638 | 0.0574 | 1.5064 | 0.0227 | 0.4118 | 0.0144 | 2.6263 | 0.0914 | 0.3635 | 0.0015 |
| 8/3/2017 | WR | 171.1748 | 0.0586 | 1.6338 | 0.0246 | 0.9905 | 0.0138 | 5.1305 | 0.0977 | 0.9854 | 0.0025 |
| 8/3/2017 | WR | 218.8899 | 0.0585 | 1.9755 | 0.0255 | 3.3881 | 0.0139 | 18.7925 | 0.1000 | 3.8346 | 0.0064 |
| 9/5/2017 | Msk. | 50.2515 | 0.0296 | 0.4393 | 0.0274 | 0.1086 | 0.0174 | 0.8244 | 0.2110 | 0.0928 | 0.0007 |
| 9/5/2017 | Msk. | 217.9942 | 0.0366 | 2.0657 | 0.0283 | 0.4534 | 0.0172 | 3.3504 | 0.2107 | 0.4456 | 0.0014 |
| 9/5/2017 | Msk. | 315.1961 | 0.0431 | 3.0475 | 0.0272 | 0.5556 | 0.0159 | 4.7549 | 0.2177 | 0.4979 | 0.0016 |
| 9/5/2017 | Msk. | 413.5451 | 0.0466 | 3.9813 | 0.0272 | 0.7026 | 0.0167 | 6.2227 | 0.2312 | 0.5915 | 0.0018 |
| 9/5/2017 | Msk. | 481.8547 | 0.0577 | 4.6160 | 0.0268 | 0.8425 | 0.0159 | 6.8466 | 0.2094 | 0.6558 | 0.0017 |
| 9/5/2017 | Msk. | 959.8993 | 0.0684 | 9.1878 | 0.0266 | 1.7937 | 0.0168 | 14.3373 | 0.2298 | 1.3877 | 0.0024 |
| 9/5/2017 | Msk. | 1280.4240 | 0.1215 | 12.1554 | 0.0262 | 2.5405 | 0.0163 | 20.8808 | 0.2241 | 1.9119 | 0.0030 |
| 9/5/2017 | Msk. | 899.6876 | 0.0844 | 8.6141 | 0.0275 | 1.9631 | 0.0170 | 16.6586 | 0.2236 | 1.4643 | 0.0028 |
| 9/5/2017 | Msk. | 558.2383 | 0.0548 | 5.3040 | 0.0257 | 1.5136 | 0.0159 | 13.3282 | 0.2291 | 1.1743 | 0.0023 |
| 9/5/2017 | Msk. | 513.4420 | 0.0501 | 4.8578 | 0.0272 | 1.8066 | 0.0160 | 18.6144 | 0.2281 | 1.3609 | 0.0025 |
| 9/5/2017 | Msk. | 688.4689 | 0.0689 | 6.5720 | 0.0269 | 3.5894 | 0.0157 | 42.8614 | 0.2030 | 2.6494 | 0.0036 |
| 9/5/2017 | Msk. | 470.7229 | 0.0555 | 4.4235 | 0.0272 | 2.0250 | 0.0162 | 21.8962 | 0.2166 | 1.4832 | 0.0027 |
| 9/5/2017 | Msk. | 117.3431 | 0.0338 | 1.0823 | 0.0261 | 0.4082 | 0.0164 | 4.3611 | 0.2166 | 0.2758 | 0.0012 |

**Notes**

Isotope intensities in fA

Intensities corrected for baseline, blank, detector calibration and decay.

Mass spectrometer sensitivity: 4x10-16 mol/fA

Uncertainties reported at 1 sigma

Trapped initial argon ratios
(40Ar/36Ar)$_o$          0
(40Ar/38Ar)$_o$          0

Minor irradiation parameters
(38Ar/37Ar)Ca     1.96E-5 ± 8.160000E-7
(37Ar/39Ar)K      1.4E-4 ± 7E-6
P(36Cl/38Cl)      250

--- Decay constants ---
λ $^{40}$K total       5.463E-10
λ $^{37}$Ar            0.01975
λ $^{39}$Ar            7.07E-06
λ $^{36}$Cl            6.308E-09

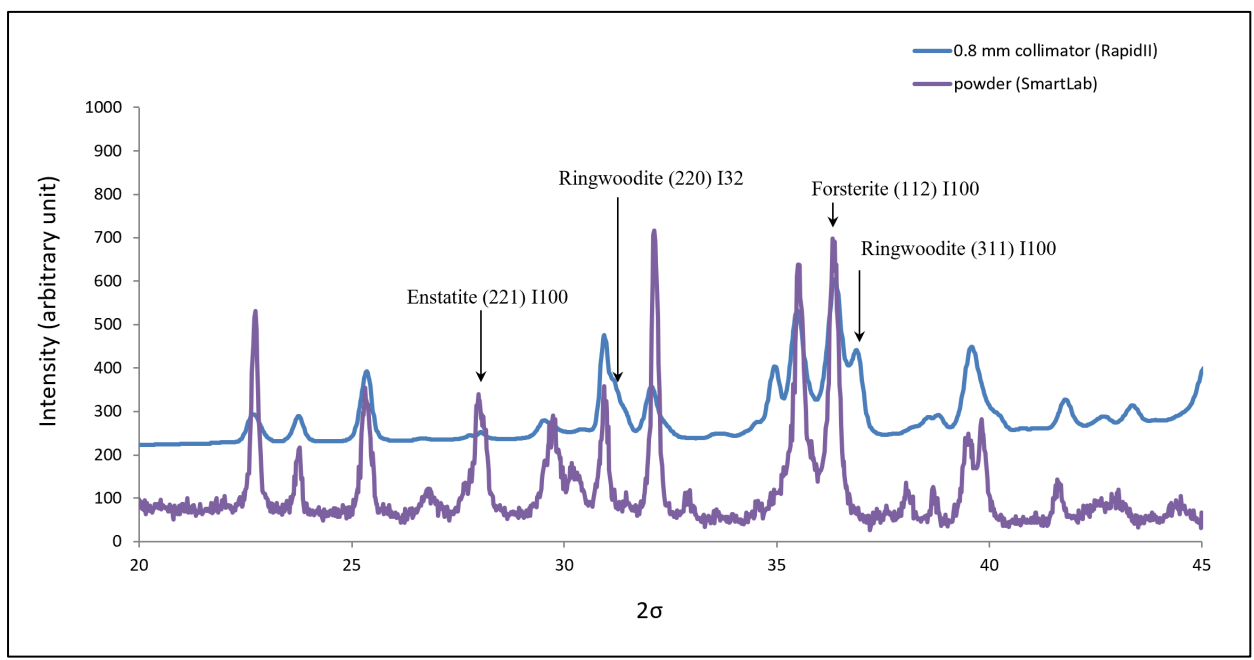

**Fig. A1** X-ray diffraction intensity data for whole rock powder and 0.8 mm collimated analysis of a melt pocket containing ringwoodite in NWA 11042. All remaining unlabeled peaks can be indexed as belonging to enstatite or forsterite.

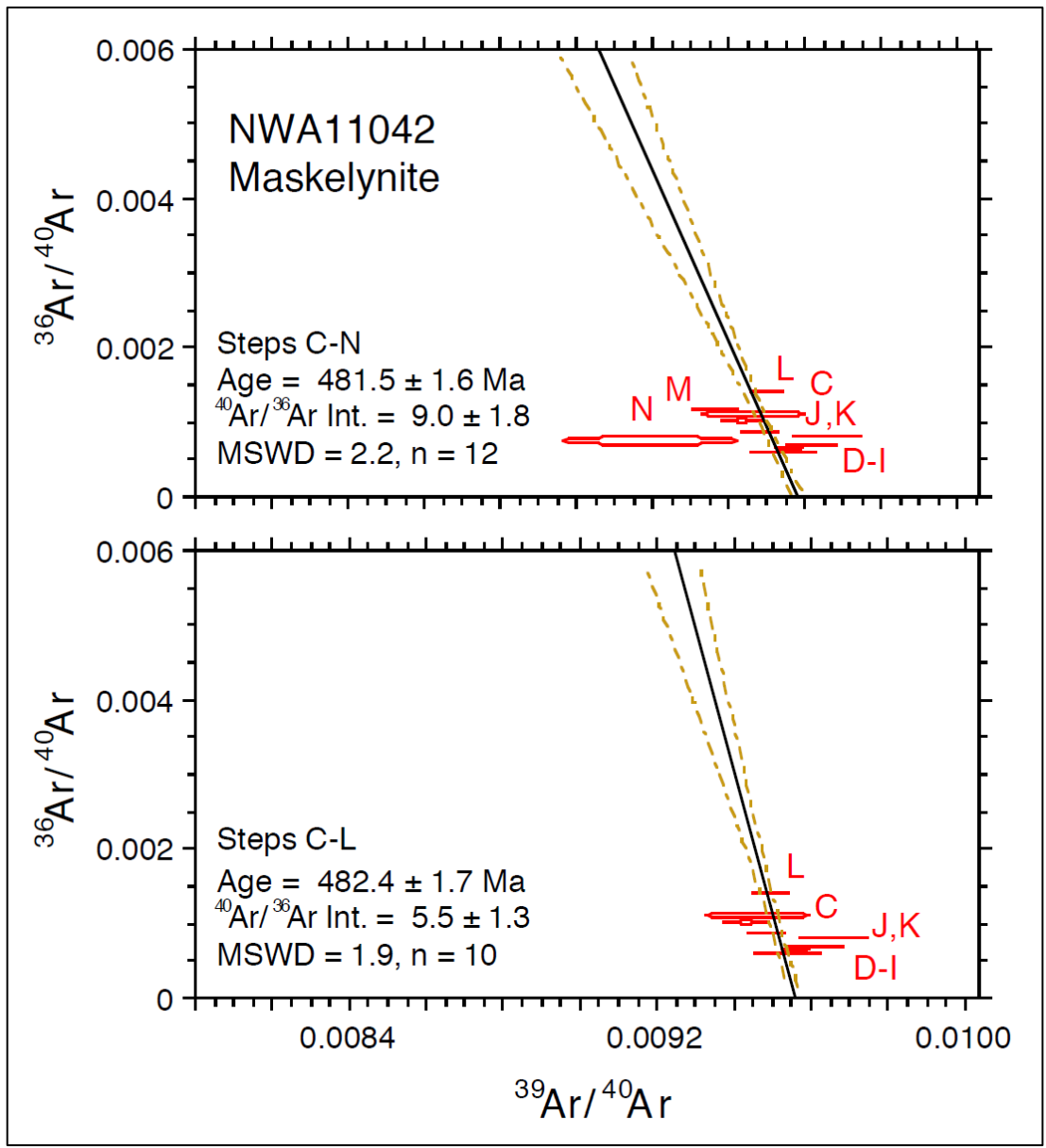

**Fig. A2** Isochron diagram of maskelynite laser heating steps C through N (top) and C through L (bottom), corrected for cosmogenic $^{36}$Ar. The minimal age difference between the isochron ages and the plateau age suggests a low to non-existent non-radiogenic component within the maskelynite.